\documentclass[namedreferences]{SolarPhysics}
\usepackage[optionalrh]{spr-sola-addons} % For Solar Physics
\usepackage{graphicx}        % For eps figures, newer & more powerfull
\usepackage{color}           % For color text: \color command
\usepackage{url}             % For breaking URLs easily trough lines
\newcommand{\etal}{{\it et al.}}

\newcommand{\aap}{    {\it Astron. Astrophys.}}

\newcommand{\apj}{    {\it Astrophys. J.}}

\newcommand{\pasj}{   {\it Pub. Astron. Soc. Japan}}

\newcommand{\solphys}{{\it Solar Phys.}}
\newcommand{\sovast}{ {\it Sov. Astron.}}

\begin{document}

\begin{article}

\begin{opening}

\title{Multilevel Analysis of Oscillation Motions in Active Regions of the Sun}

\author{V.E.~\surname{Abramov-Maximov}$^{1}$\sep
        G.B.~\surname{Gelfreikh}$^{1}$\sep
        N.I.~\surname{Kobanov}$^{2}$\sep
        K.~\surname{Shibasaki}$^{3}$\sep
        S.A.~\surname{Chupin}$^{2}$
       }
\runningauthor{V.E.~Abramov-Maximov \etal}
\runningtitle{Multilevel Analysis of Oscillation Motions in Active
Regions of the Sun}

   \institute{$^{1}$ Central Astronomical Observatory at Pulkovo, Russian Acad.
   Sci., St.Petersburg, 196140, Russia\\
                     email: \url{beam@gao.spb.ru}\\
              $^{2}$ Institute of Solar-Terrestrial Physics,
P.O. Box 4026, Irkutsk, 664033, Russia\\
                     email: \url{kobanov@iszf.irk.ru} \\
                     $^{3}$ Nobeyama Solar Radio Observatory, Minamisaku, Nagano,
                     Japan\\
                     email: \url{shibasaki@nro.nao.ac.jp} \\
             }

\begin{abstract}
The nature of  the three-minute and five-minute oscillations
observed in sunspots is considered to be an effect of propagation
of magnetohydrodynamic (MHD) waves from the photosphere to the
solar corona. However, the real modes of these waves and the
nature of the filters that result in rather narrow frequency bands
of these modes are still far from being generally accepted in
spite of a large amount of observational material obtained  in a
wide range of wave bands of observations. The significance of this
field of research is based on the hope that local seismology can
be used to find the structure of the solar atmosphere in magnetic
tubes of sunspots. We expect that substantial progress can be
achieved by simultaneous observations of the sunspot oscillations
in different layers of the solar atmosphere in order to gain
information on propagating waves. In this study we used a new
method that combines the results of an oscillation study made in
optical and radio observations. The optical spectral measurements
in photospheric and chromospheric lines of the line-of-sight
velocity were carried out at the Sayan Solar Observatory. The
radio maps of the Sun were obtained with the Nobeyama
Radioheliograph at 1.76~cm. Radio sources associated with the
sunspots were analyzed to study the oscillation processes in the
chromosphere--corona transition region in the layer with magnetic
field $B=2000$~G. A high level of instability of the oscillations
in the optical and radio data was found. We used a wavelet
analysis for the spectra. The best similarities of the spectra of
oscillations obtained by the two methods were detected in the
three-minute oscillations inside the sunspot umbra for the dates
when the active regions were situated near the center of the solar
disk. A comparison of the wavelet spectra for optical and radio
observations showed a time delay of about 50~seconds of the radio
results with respect to optical ones. This implies a MHD wave
traveling upward inside the umbral magnetic tube of the sunspot.
For the five-minute oscillations the similarity in spectral
details could be found only for optical oscillations at the
chromospheric level in the umbra region or very close to it. The
time delays seem to be similar. Besides three-minute and
five-minute ones, oscillations with longer periods (8 and 15
minutes) were detected in optical and radio records. Their nature
still requires further observational and theoretical study though
for even a preliminary discussion.

\end{abstract}
\end{opening}
%-------------------------------------------------

\section{Introduction}
     \label{S-Introduction}

The study of oscillatory motions is one of the most significant
methods of plasma structure analysis of the solar
atmosphere~(Beckers and Schultz, 1972; Bhatnagar, Livingston and
Harvey, 1972; Zhugzhda and Dzhalilov, 1982; Staude, 1999;
Zhugzhda, Balthasar, and Staude,
2000). %\cite{Beckers72,Bhatnagar72,Zhugzhda82,Staude99,Zhugzhda00}.
Currently investigations are carried out in different wavelength
ranges from radio to EUV and X-rays, which are based on ground and
cosmic observations. Because the structuring of the plasma is
directly connected with such fundamental astrophysical problems as
accumulation and release of energy, its investigation with this
quickly developing method is essential for progress in
understanding the physics of coronal heating and the origin of
flares. Quasi-periodic oscillations~(QPO) are recorded in
practically all wavelength ranges and  in all structures of the
solar atmosphere and show up at isolated frequencies with periods
from fractions of a minute to hours or even
days~\cite{Bogdan00,Fludra01,Gelfreikh06}. Most of these
oscillations are of an unstable nature: both amplitude and
frequency vary with time and the oscillations are often seen as
packets of about a dozen periods. Wavelet analysis is a more
appropriate tool than the traditional spectral Fourier analysis.

The observed QPOs are caused by magneto-hydrodynamic waves of
different kinds. They include oscillations of the local closed
resonance structures as well as global ones, and effects of MHD
wave propagation of different types. In the latter case an
explanation for the appearance of some isolated periods requires
filtering on the way to the region where the observed emission is
generated. The choice of an adequate theory for interpreting these
observations includes several essentially complicated problems
that do not have an adequate solution at present, and involves
models of both the solar atmosphere and the layers below the solar
surface.

Though many types of QPOs, especially above sunspots, have been
studied for decades~(Lites, 1992; Lites et al., 1998; Bogdan,
2000; Bogdan and Judge,
2006), %\cite{Lites92,Lites98,Bogdan00,Bogdan06},
their nature and fine structure are not yet satisfactorily
understood. One of the reasons is directly related to the problems
of satisfactory and reliable methods of observations. For many
years the investigations of the sunspot-associated oscillations
were conducted only by optical methods that represented the
phenomena at the level of the photosphere and chromosphere. Modern
techniques of observations from space facilitate the study of the
oscillations in the corona and transition region. However, the
magnetic tube of a sunspot does not appear in these pictures as it
could be if followed on a lower level of optical identifications.
Progress in those studies was achieved with radio mapping of the
sun at microwaves. On the radio maps the sunspots are clearly seen
as bright highly polarized small radio sources generated by
thermal cyclotron emission of coronal electrons at lower harmonics
of their gyrofrequency; magnetic field of thousands of Gauss is
needed for the effect~\cite{Gelfreikh79,Nindos96,Nindos02}.

For the analysis of the three-dimensional structure of the QPOs
one needs parallel observations in several wavelength ranges. For
a realization of this program, however, it is necessary to make
simultaneous observations of same object. New results of such
observational programs are the main topic of the present paper.

We have analyzed the active region NOAA~10661 in August~2004 from
the Sayan Solar Observatory~\cite{Kobanov06} and simultaneous
radio measurements of the same AR at wavelength $\lambda=1.76$~cm
found from radio maps of the Nobeyama Radioheliograph (similar to
\opencite{Gelfreikh99}; \opencite{Shibasaki01a};
\opencite{Shibasaki01b}). The magnetic field at the photosphere
was measured in the line Fe\,{\sc i}~6569~\AA{} and
line-of-sight~(LOS) velocities in the chromosphere registered in
H$\alpha$ line. Radio maps with 10\,--\,15 arcsec resolution, both
in intensity and circular polarization, clearly showed the
structure of the analyzed AR with compact bright sources above
large sunspots. The comparison of observational data was carried
out with a wavelet spectral analysis made for sunspot details and
other structures registered in optical observations and
sunspot-associated brightness found from the radio maps.

Both optical and radio wavelet spectra show many identical
structures. The differences are also discussed. The degree of
similarity of the two types of observation vary with the position
of the AR on the solar disk. Future directions of the development
of the illustrated method of the solar physics are discussed.

\section{Optical Observations} \label{S-optical_observations}

Optical observations were carried out with the horizontal solar
telescope of the Sayan Solar Observatory of the Institute of
Solar-Terrestrial Physics of the Siberian Division of the Russian
Academy of Sciences~(Kobanov and Makarchik, 2004a; Kobanov and
Makarchik, 2004b). %\cite{Kobanov04a,Kobanov04b}.
The line-of-sight~(LOS) velocity and longitude component of the
magnetic field (using line Fe\,{\sc i}~6559~\AA, g=1.4) at the
photosphere and the LOS velocity in the chromospheric H$\alpha$
line were obtained. The slit of the spectrograph was
east\,--\,west directed, crossing the center of the sunspot. The
spatial resolution was about one arcsec along the spectrograph
slit. The instrumental shifts of the spectrum were determined
using an H$_2$O telluric line near H$\alpha$. These shifts were
subtracted from the calculated signals. In order to eliminate
aliasing, the time between expositions  was shorter than the
duration of an exposition.

We used a Dove prism installed just in front of the slit. The
prism was usually rotated so that the slit was parallel to the
east-west direction on the sunspot image.

The duration of one set of observations was about one hour with a
cadence of a few seconds. While analyzing the observations we used
wavelet spectra with Morlet functions of the sixth order as a base
function~(see \opencite{Torrence98}, \opencite{Kobanov06}). We use
the 95\% confidence level (2 $\sigma$) as criterion of reliability
of the oscillations.

Table~\ref{T-ar} presents the main parameters of the optical
observations used in this study.

\begin{table} \caption{Coordinates of the active region AR661
and intervals of optical observations.}
\begin{tabular}{ccc}
\hline
Date & Coordinates & Time of observations \\
     & & UT \\
\hline
2004--Aug--15 & (N 07,E 58)  & 04:04:00 -- 05:04:00 \\
2004--Aug--16 & (N 07,E 46)  & 00:58:00 -- 01:41:00 \\
2004--Aug--18 & (N 07,E 20)  & 01:01:00 -- 01:43:00 \\
\hline
\end{tabular}
\label{T-ar}
\end{table}

\section{Radio Observations}
\label{S-radio_observations}

Nobeyama Radioheliograph~(Nakajima et al., 1994) %\cite{Nakajima94}
observations at 1.76~cm were used to study the same sunspots.
Radio maps of the whole disk were constructed for the dates of
observations with the temporal interval of ten seconds between the
maps and ten seconds averaging. We constructed intensity and
circular polarization maps. Bearing in mind that the
sunspot-associated sources generated by thermal gyroresonance
emission at the third harmonic of the electron gyrofrequency are
the most sensitive to oscillation processes, we studied only those
cases where such sources were present. The identification of the
nature of a source is based on its high brightness temperature and
strong polarization. The 2D spatial resolution of the radio maps
was about 10\,--\,15 arcsec.

For the wavelet analysis we used the same method as for the
analysis of optical data. We used wavelet spectra with Morlet
functions of the sixth order and the 95\% confidence level as
criterion of reliability of the oscillations. Wave-trains of
three-minute oscillations lasted more than 10 minutes. It is an
additional argument for the reality of the oscillations. Moreover,
we calculated the cross-wavelet transform and wavelet coherency
between optical and radio time
series~(Figure~\ref{cross_wavelet}). Obviously the wave-trains of
the optical and radio data agree well. Consequently this is
certainly not a noise.

\section{Comparison of Oscillations in Radio and Optical in the Frequency
Range of 3\,--\,5~mHz} \label{S-comparison}

We used simultaneous observations of the temporal variations of
sunspots in the optical and radio wavelength range covering
periods from one to ten minutes. We measured the chromospheric LOS
velocity by optical observations in the H$\alpha$ line. At 1.76~cm
radio wavelength the brightness of a radio source above a sunspot
measures oscillations of the temperature inside the layer with a
magnetic field of close to 2000 G. The comparison of the
oscillations at different levels of the solar atmosphere can help
us to find a physical interpretation of the observed phenomena.

\begin{figure}
   \centerline{\hspace*{0.05\textwidth}
               \includegraphics[width=0.388\textwidth,clip=]{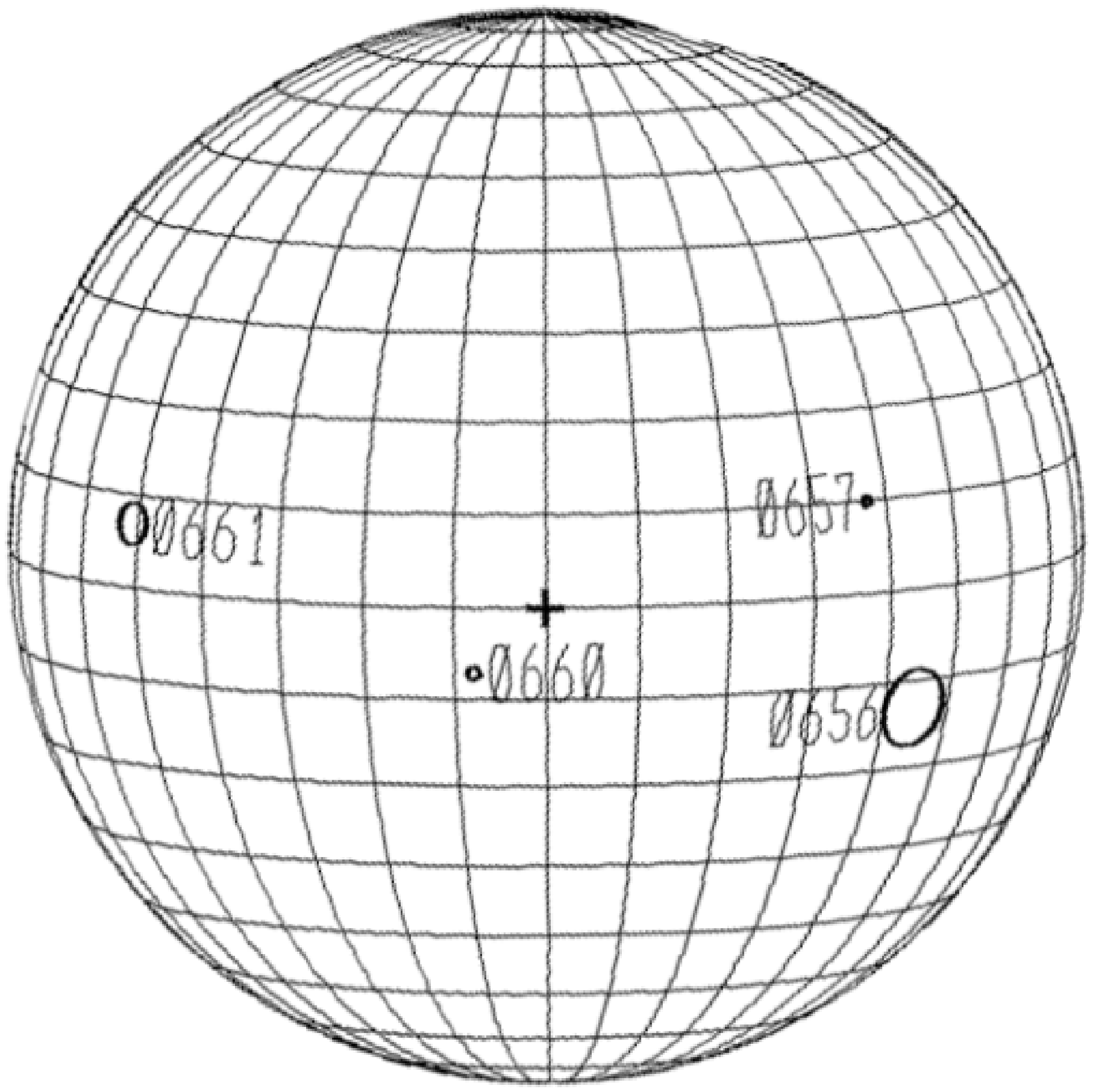}
               \hspace*{0.1\textwidth}
               \includegraphics[width=0.4\textwidth,clip=]{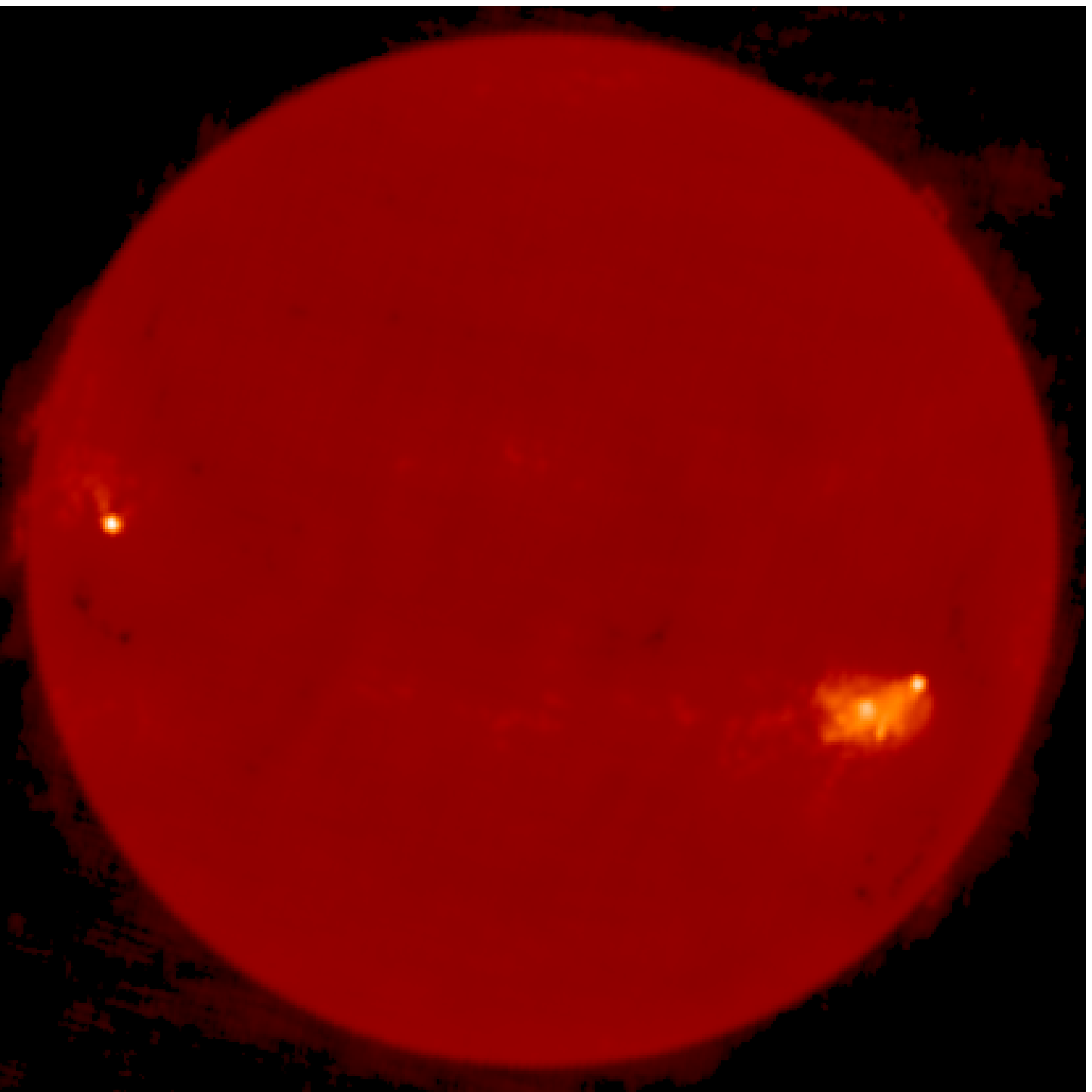}}
   \vspace{-0.38\textwidth}
   \centerline{\Large \bf
               \hspace{0.05\textwidth}\color{black}{a}
               \hspace{0.5\textwidth} \color{white}{b}
   \hfill}
   \vspace{0.38\textwidth}
   \centerline{\raisebox{0.005\textwidth}{\includegraphics[width=0.515\textwidth,clip=]{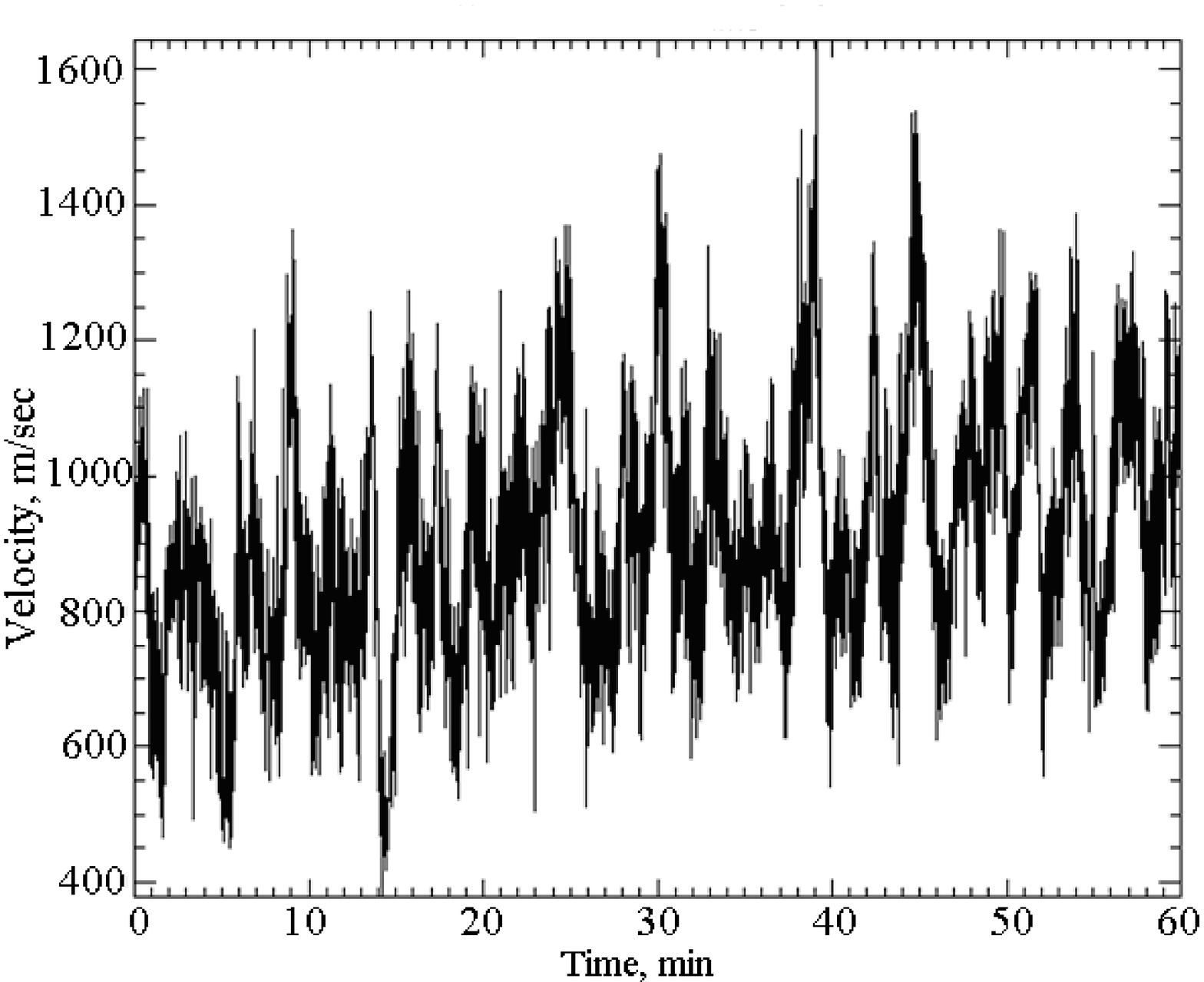}}
               \hspace*{0.02\textwidth}
               \includegraphics[width=0.515\textwidth,clip=]{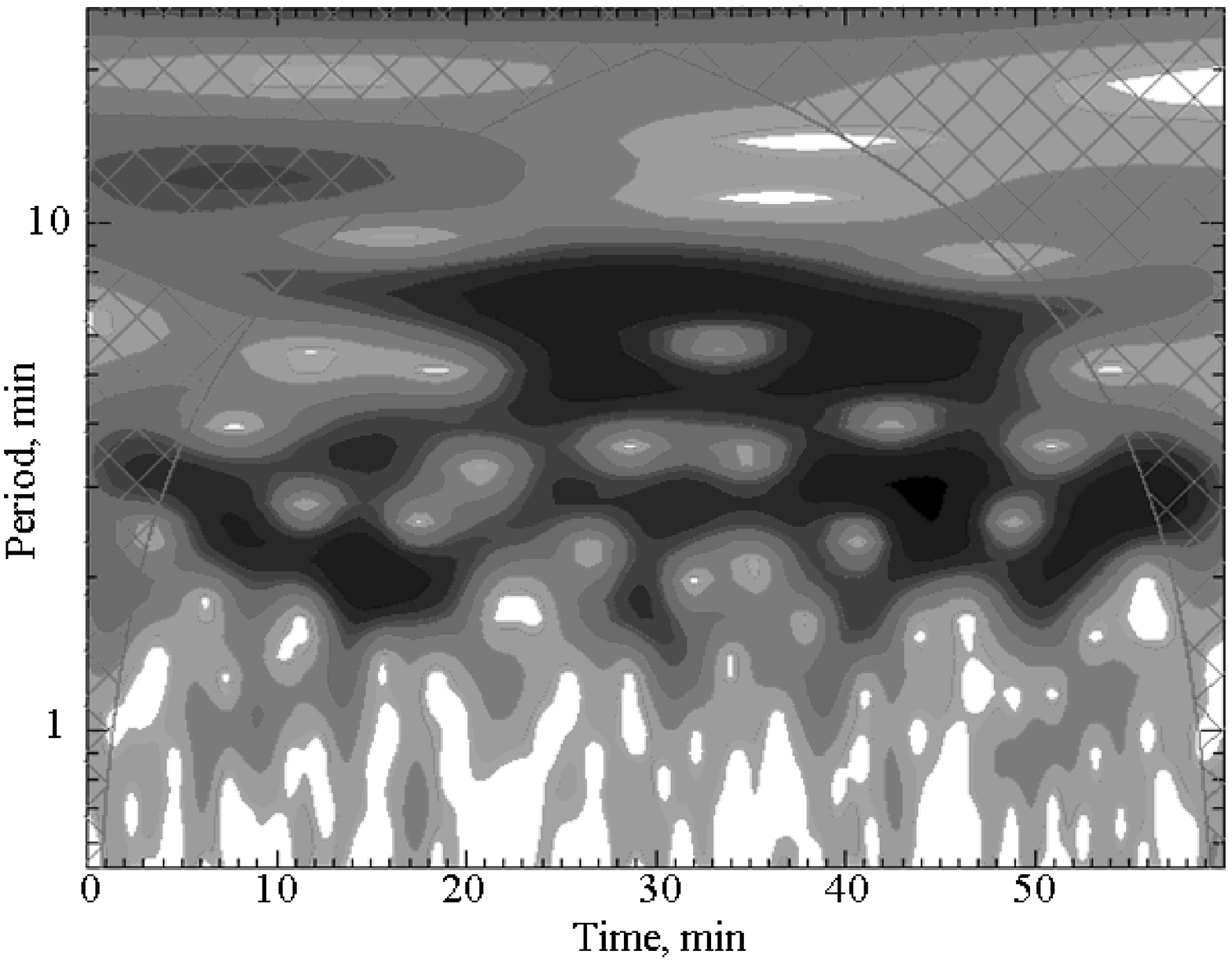}}
   \vspace{-0.39\textwidth}
   \centerline{\Large \bf
               \hspace{0.04 \textwidth} \color{black}{c}
               \hspace{0.48\textwidth} \color{white}{d}
   \hfill}
   \vspace{0.39\textwidth}
   \centerline{\includegraphics[width=0.515\textwidth,clip=]{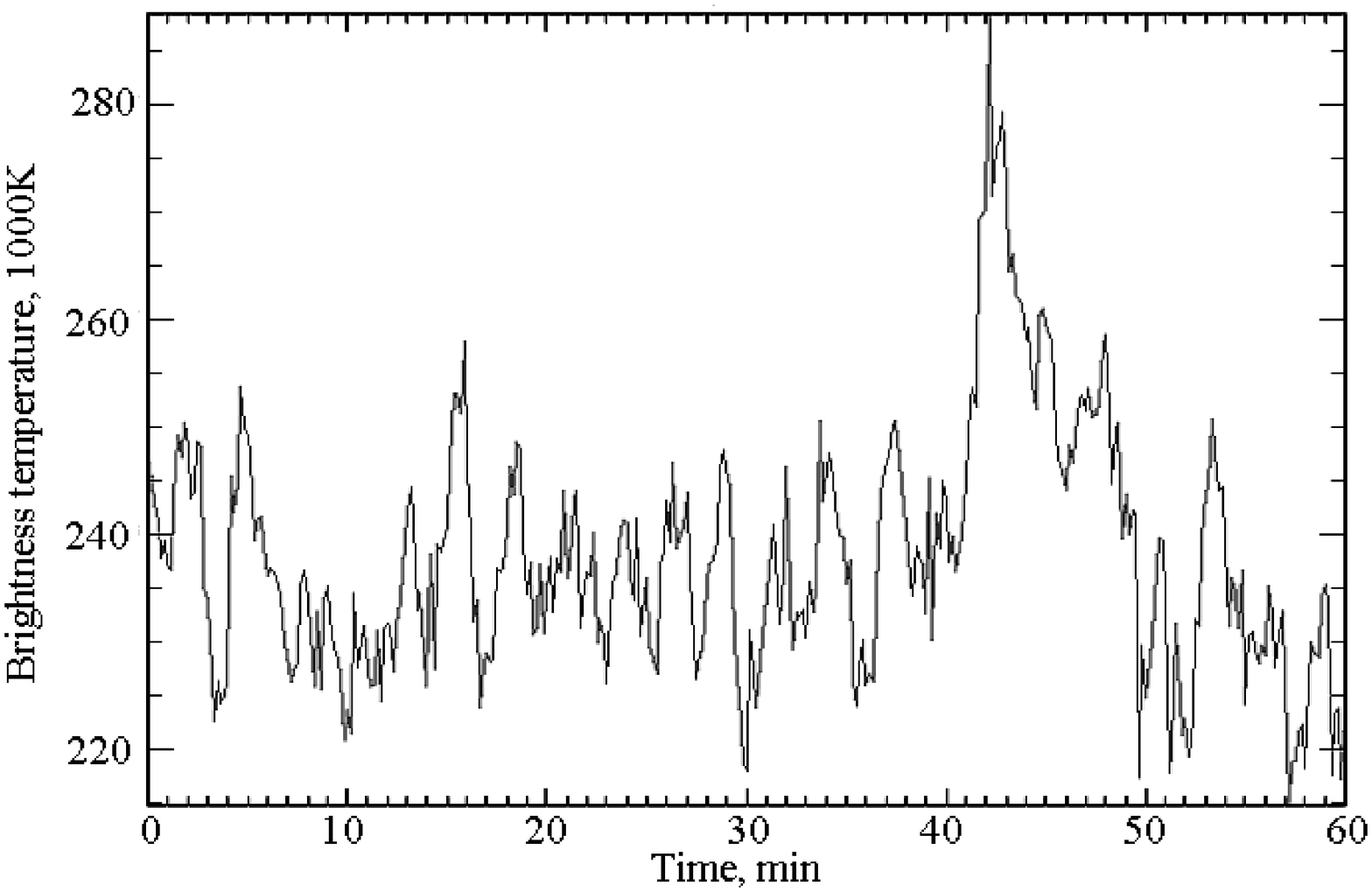}
               \raisebox{-0.002\textwidth}{\includegraphics[width=0.525\textwidth,clip=]{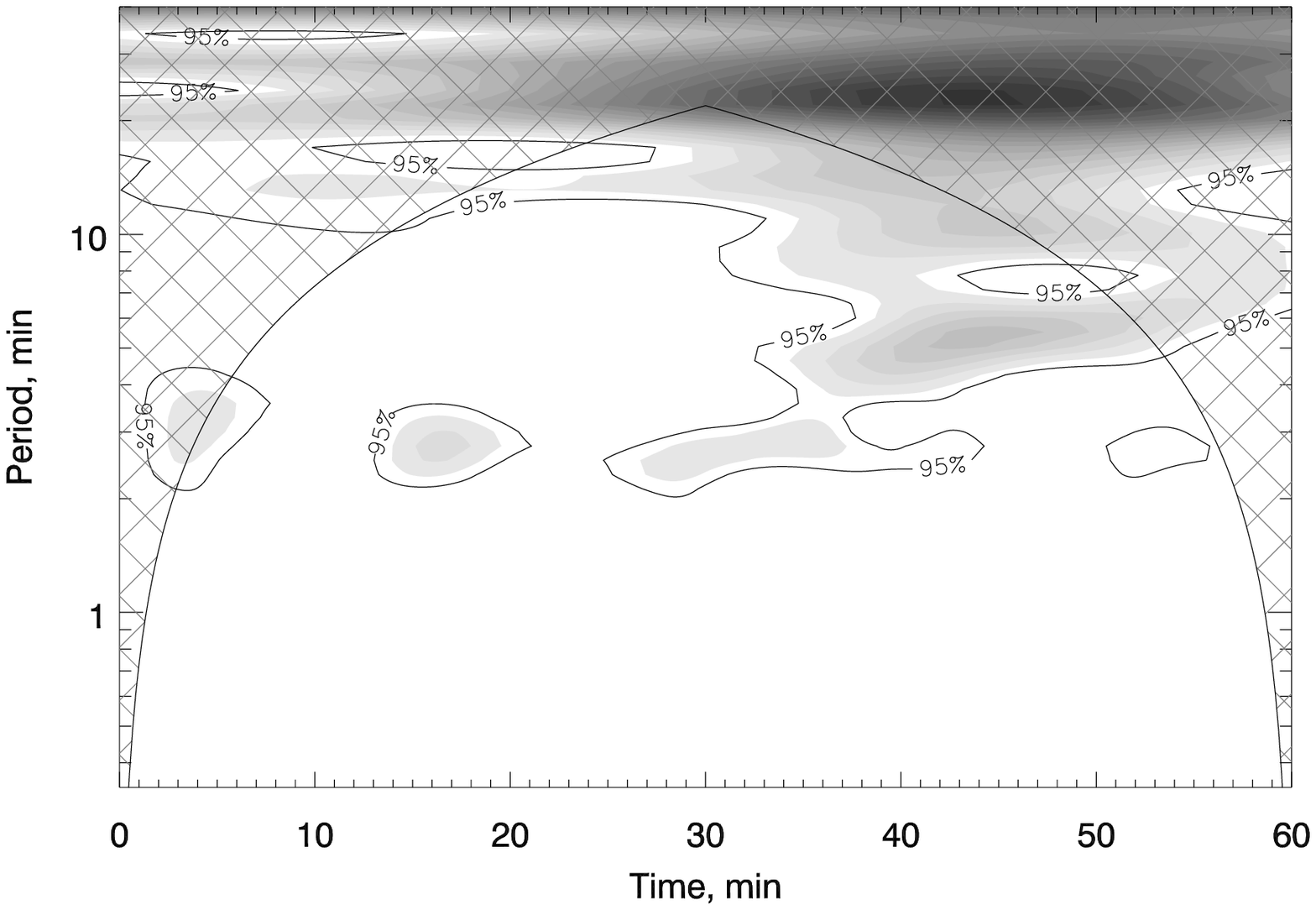}}}
   \vspace{-0.32\textwidth}
   \centerline{\Large \bf
               \hspace{0.045 \textwidth} \color{black}{e}
               \hspace{0.48\textwidth}  \color{black}{f}
   \hfill}
   \vspace{0.32\textwidth}
\caption{NOAA~661, 2004 August~15, 04:04\,--\,05:04~UT. Top: (a)
active region position on the solar disk (Mees Solar Observatory),
(b) radio map at 17~GHz. Middle: (c) the temporal variations and
(d) wavelet analysis of the chromospheric LOS velocity in
H$\alpha$. Bottom: (e) the temporal variations and (f) wavelet
analysis of the radio emission of the bright source in AR661.
Darker regions correspond to higher power. The crosshatched area
shows the cone of influence (COI). The contours show the 95\%
confidence level.} \label{F-aug15}
\end{figure}

\begin{figure}
   \centerline{\hspace*{0.05\textwidth}
               \includegraphics[width=0.388\textwidth,clip=]{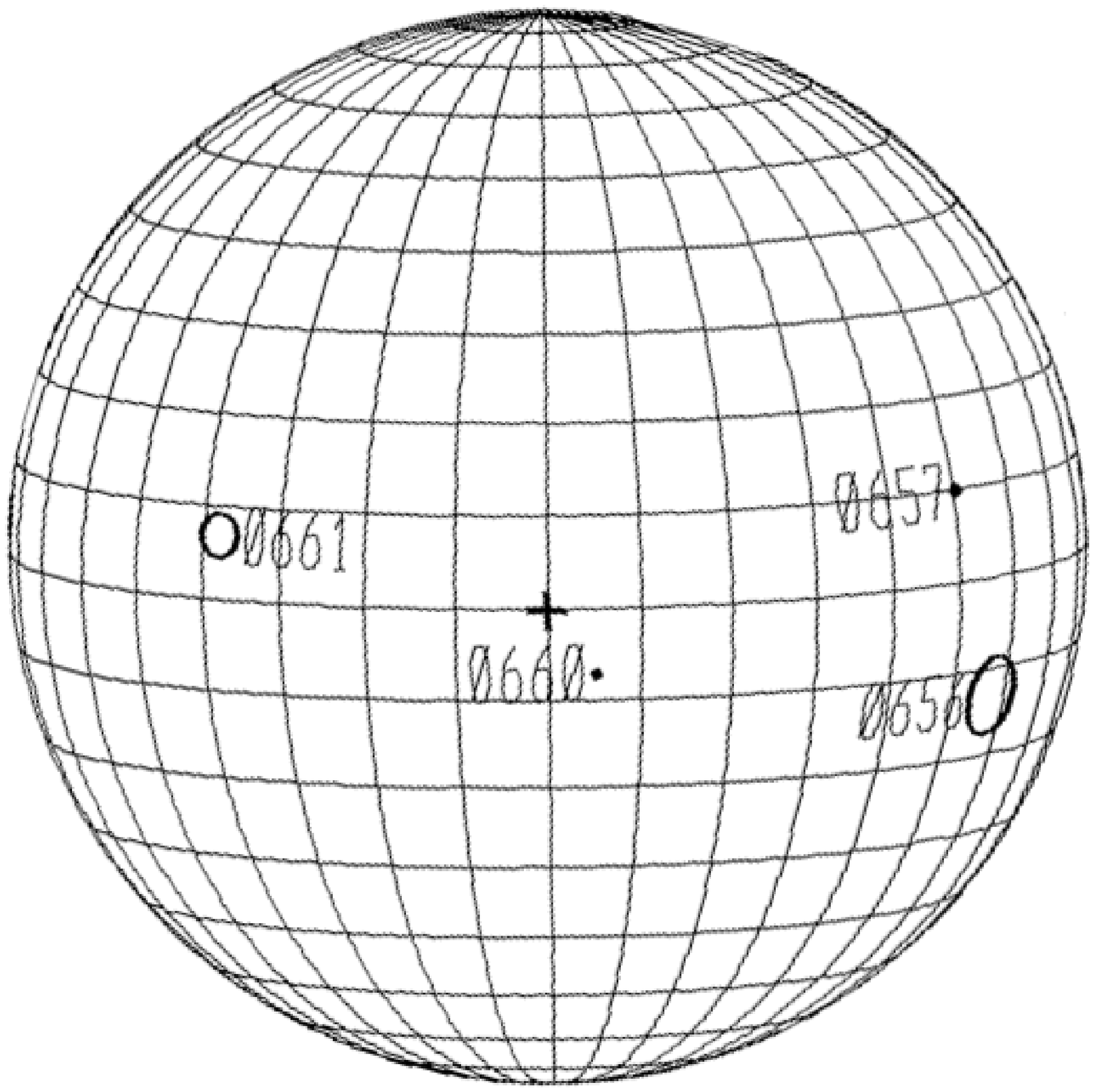}
               \hspace*{0.1\textwidth}
               \includegraphics[width=0.4\textwidth,clip=]{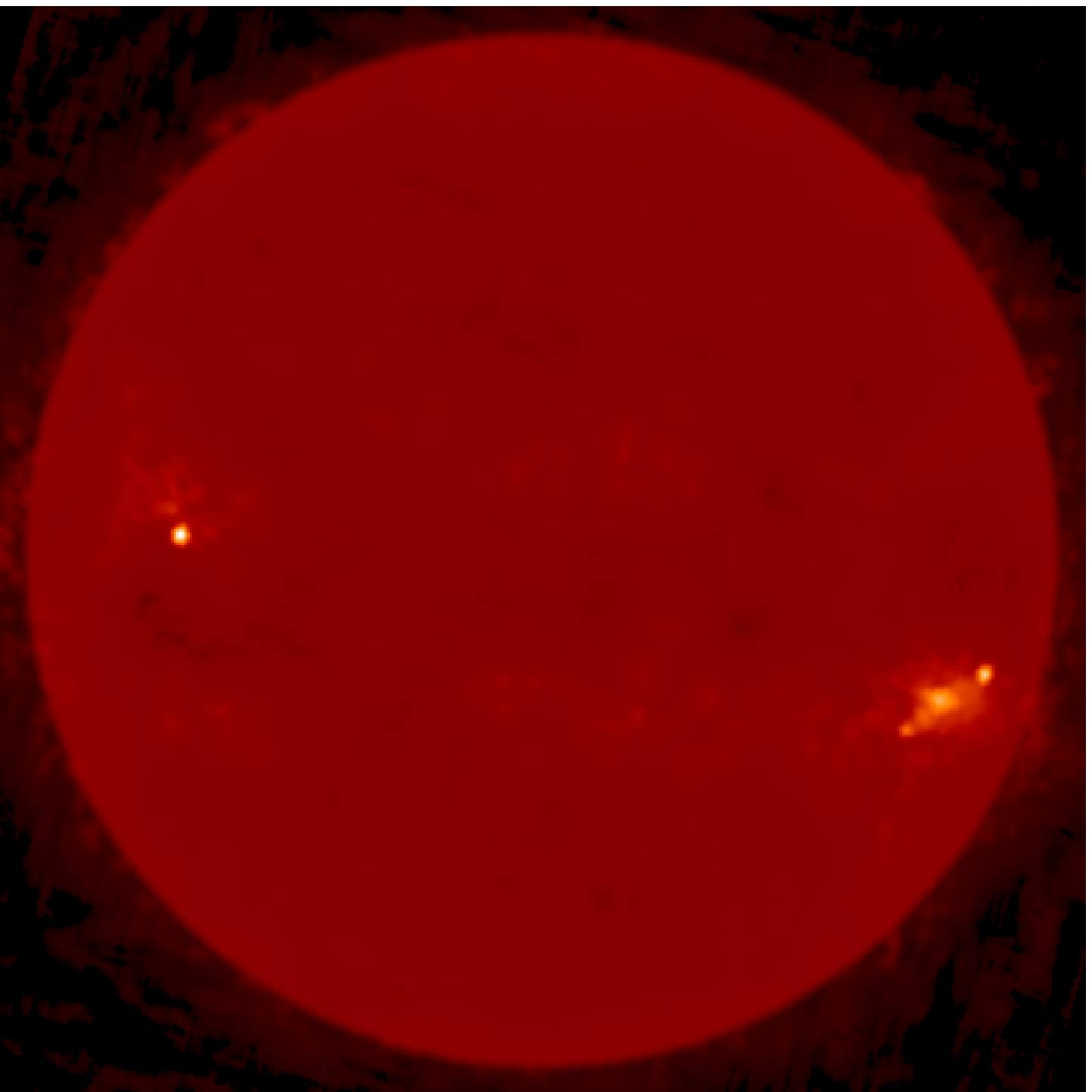}}
   \vspace{-0.36\textwidth}
   \centerline{\Large \bf
               \hspace{0.05\textwidth}\color{black}{a}
               \hspace{0.5\textwidth} \color{white}{b}
   \hfill}
   \vspace{0.36\textwidth}
   \centerline{\includegraphics[width=0.515\textwidth,clip=]{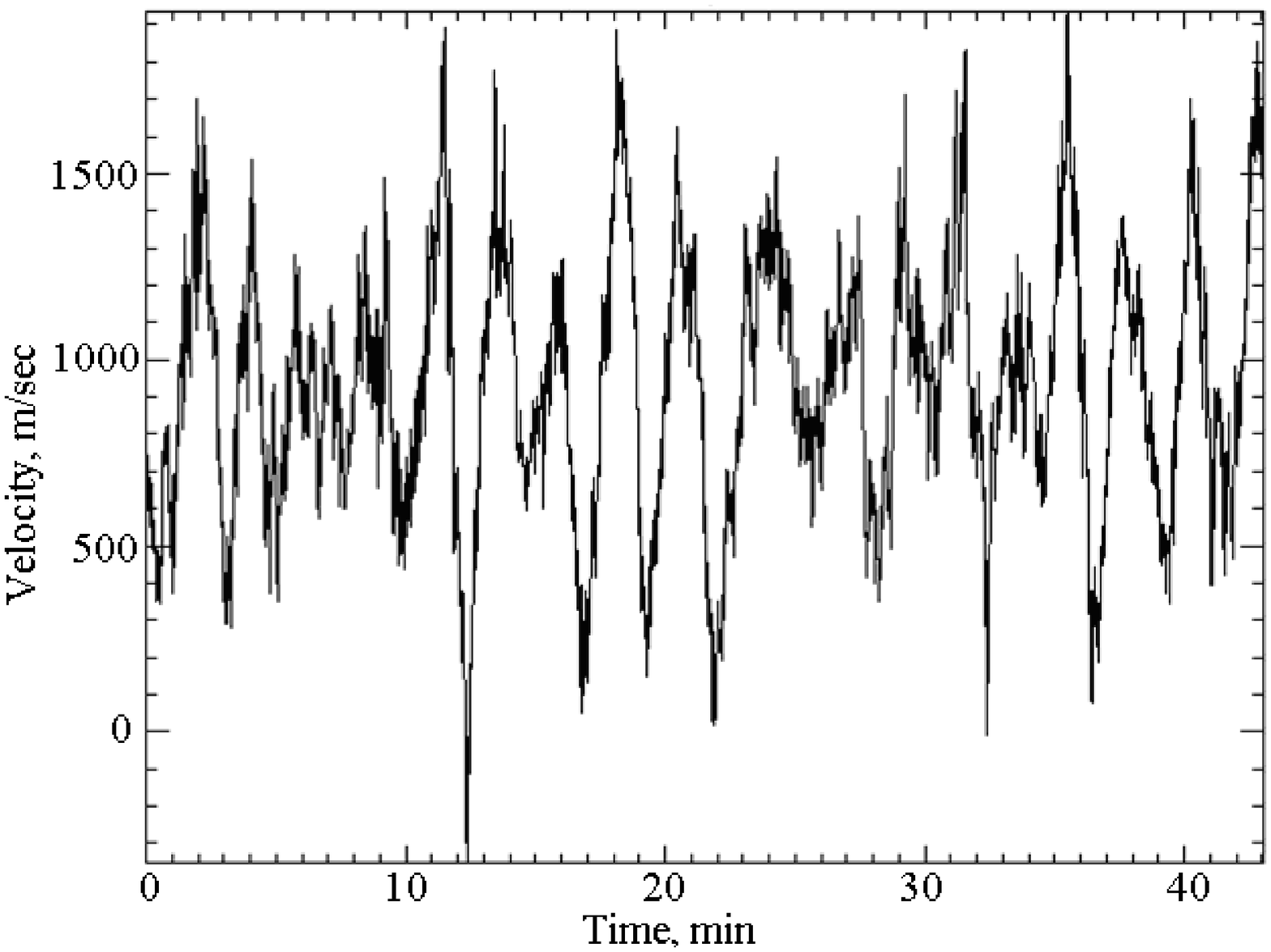}
               \hspace*{0.0\textwidth}
               \includegraphics[width=0.515\textwidth,clip=]{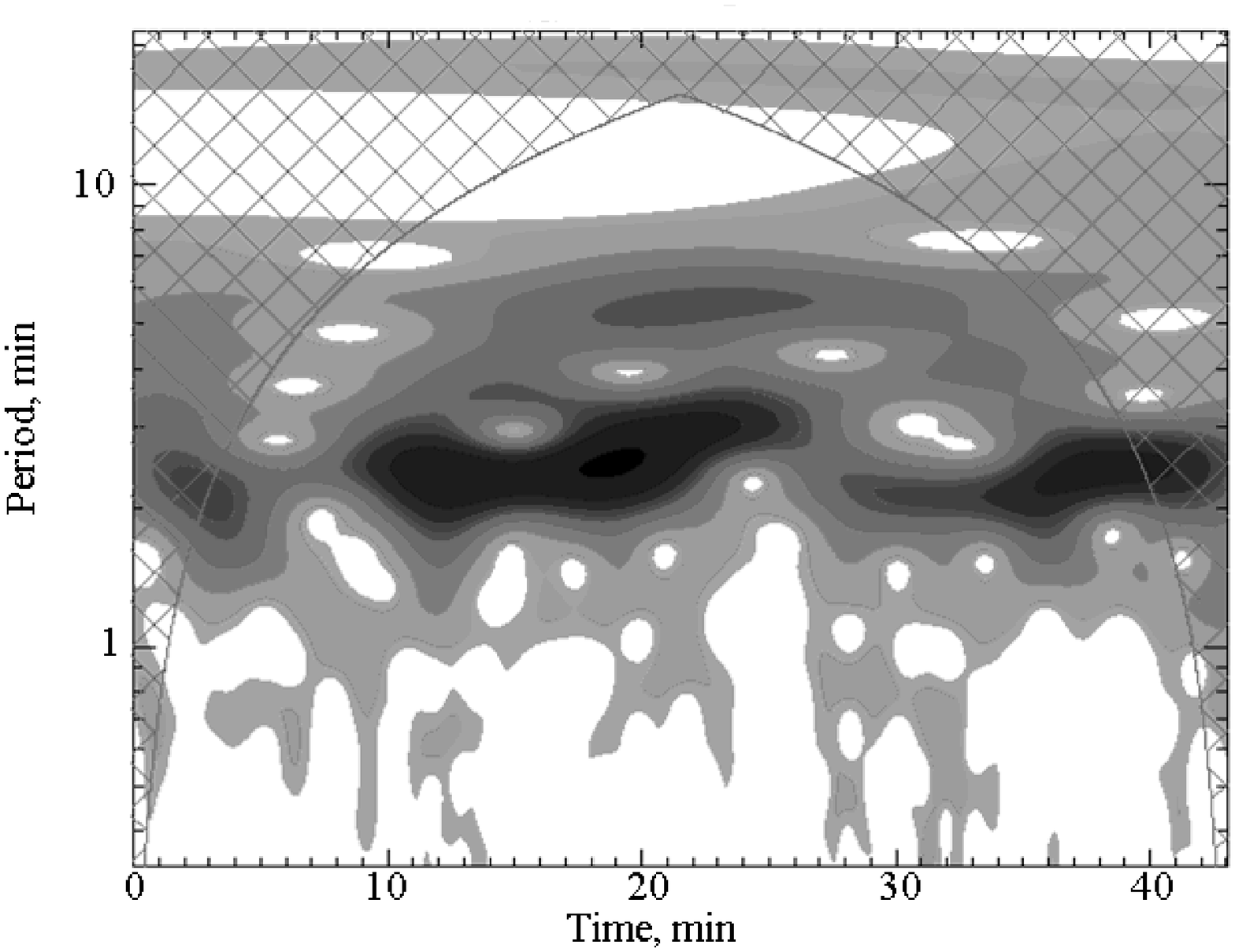}}
   \vspace{-0.37\textwidth}
   \centerline{\Large \bf
               \hspace{0.04 \textwidth} \color{black}{c}
               \hspace{0.48\textwidth} \color{black}{d}
   \hfill}
   \vspace{0.37\textwidth}
   \centerline{\includegraphics[width=0.515\textwidth,clip=]{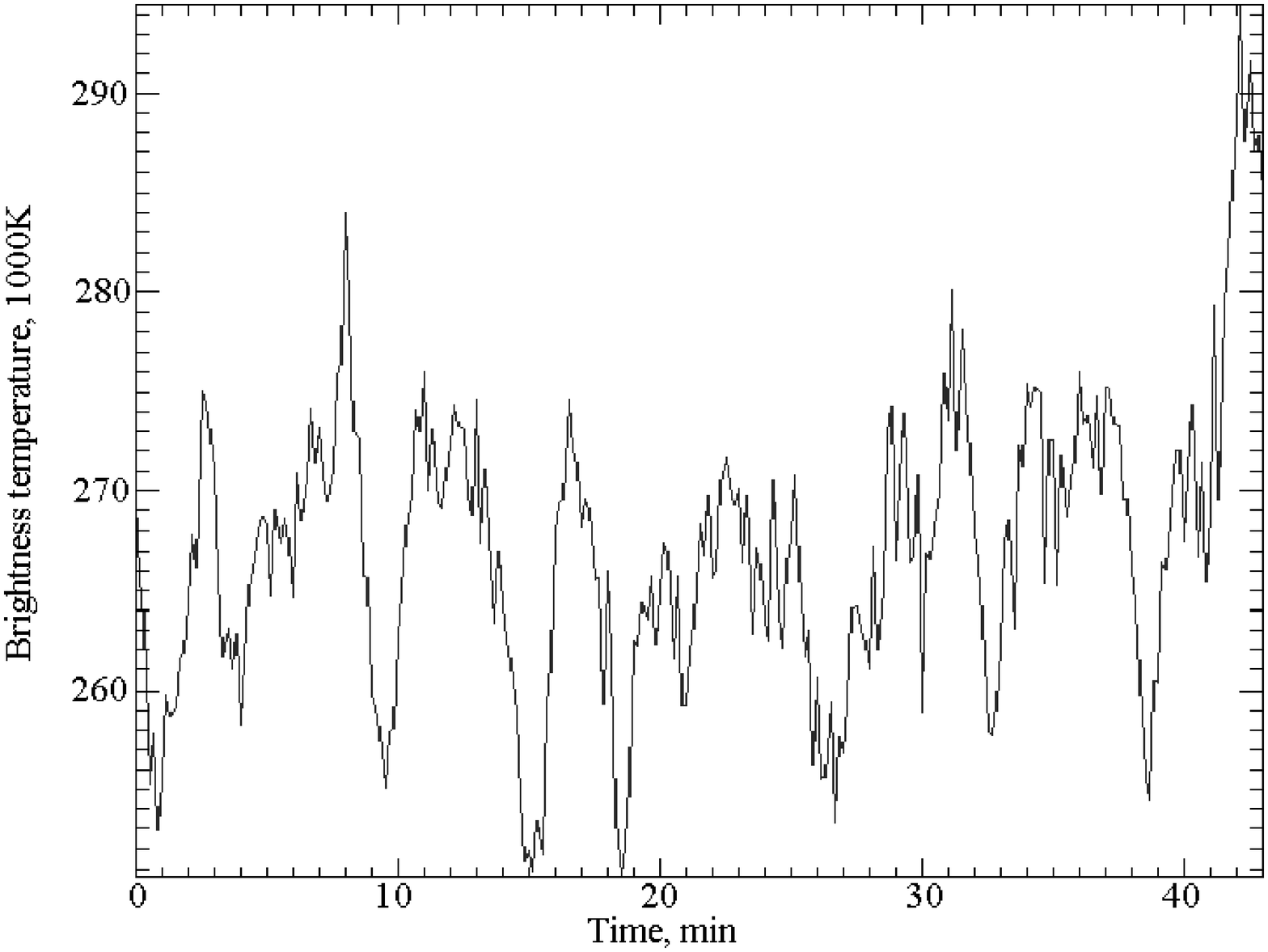}
               \includegraphics[width=0.502\textwidth,clip=]{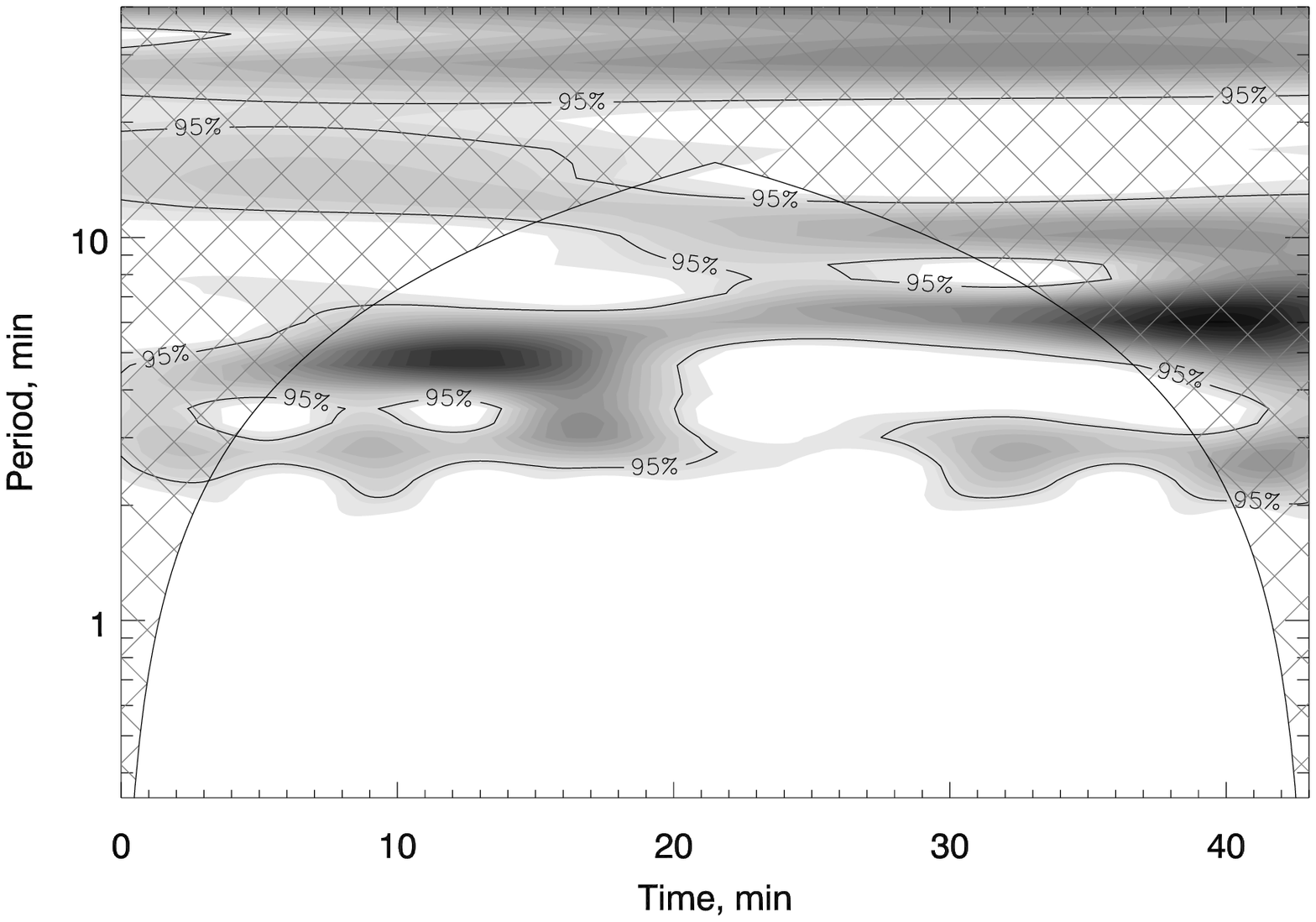}}
   \vspace{-0.35\textwidth}
   \centerline{\Large \bf
      \hspace{0.045\textwidth} \color{black}{e}
      \hspace{0.48\textwidth}  \color{black}{f}
   \hfill}
   \vspace{0.35\textwidth}

\caption{NOAA~661, 2004 August~16, 00:58\,--\,01:41 UT. Top: (a)
active region position on the solar disk (Mees Solar Observatory),
(b) radio map at 17~GHz. Middle: (c) the temporal variations and
(d) wavelet analysis of the chromospheric LOS velocity in
H$\alpha$. Bottom: (e) the temporal variations and (f) wavelet
analysis of the radio emission of the bright source in AR661.
Darker regions correspond to higher power. The crosshatched area
shows the COI. The contours show the 95\% confidence level.}
\label{F-aug16}
\end{figure}

\begin{figure}
   \centerline{\hspace*{0.05\textwidth}
               \includegraphics[width=0.388\textwidth,clip=]{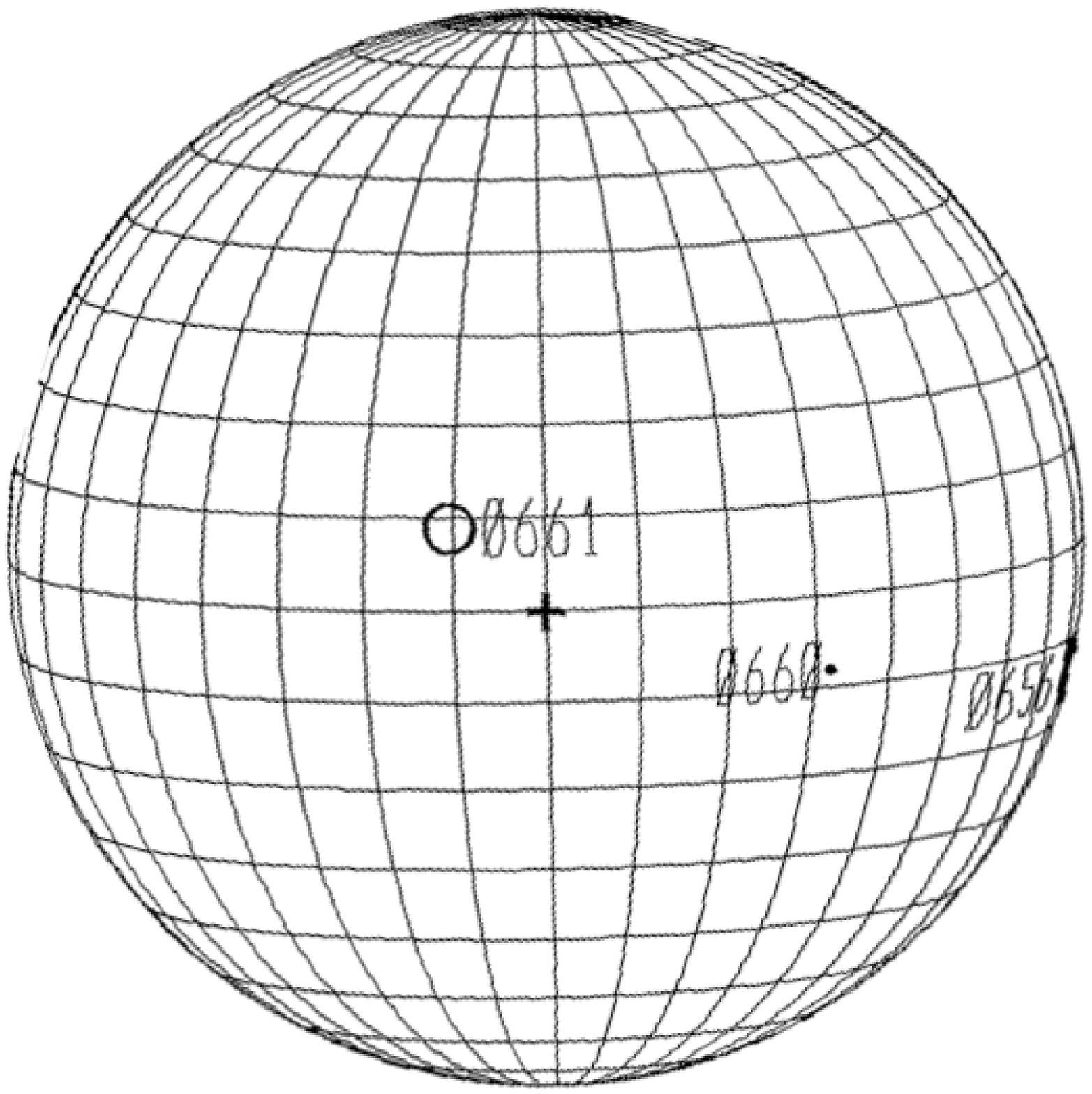}
               \hspace*{0.1\textwidth}
               \includegraphics[width=0.4\textwidth,clip=]{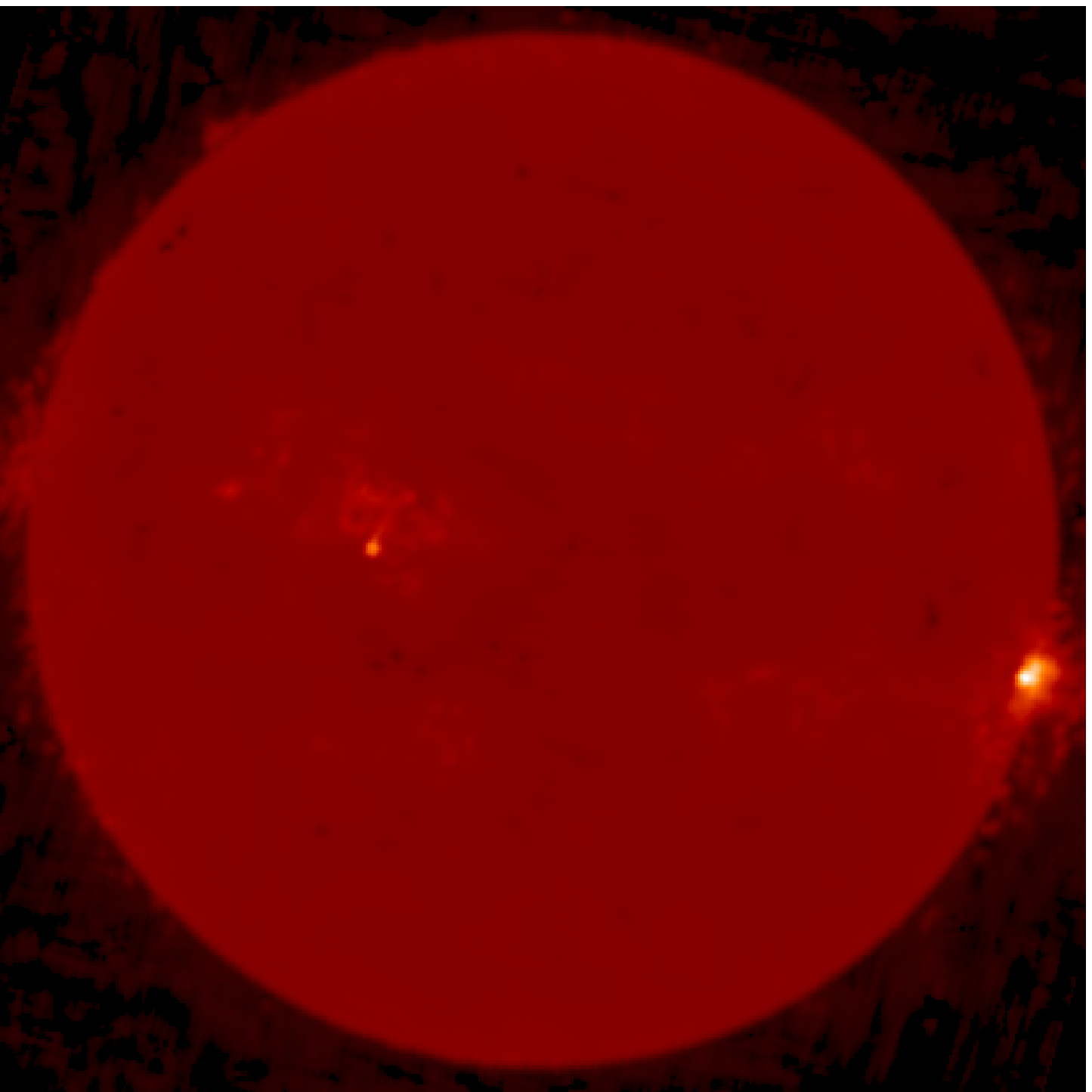}}
   \vspace{-0.36\textwidth}
   \centerline{\Large \bf
               \hspace{0.05\textwidth}\color{black}{a}
               \hspace{0.5\textwidth} \color{white}{b}
   \hfill}
   \vspace{0.36\textwidth}
   \centerline{\includegraphics[width=0.515\textwidth,clip=]{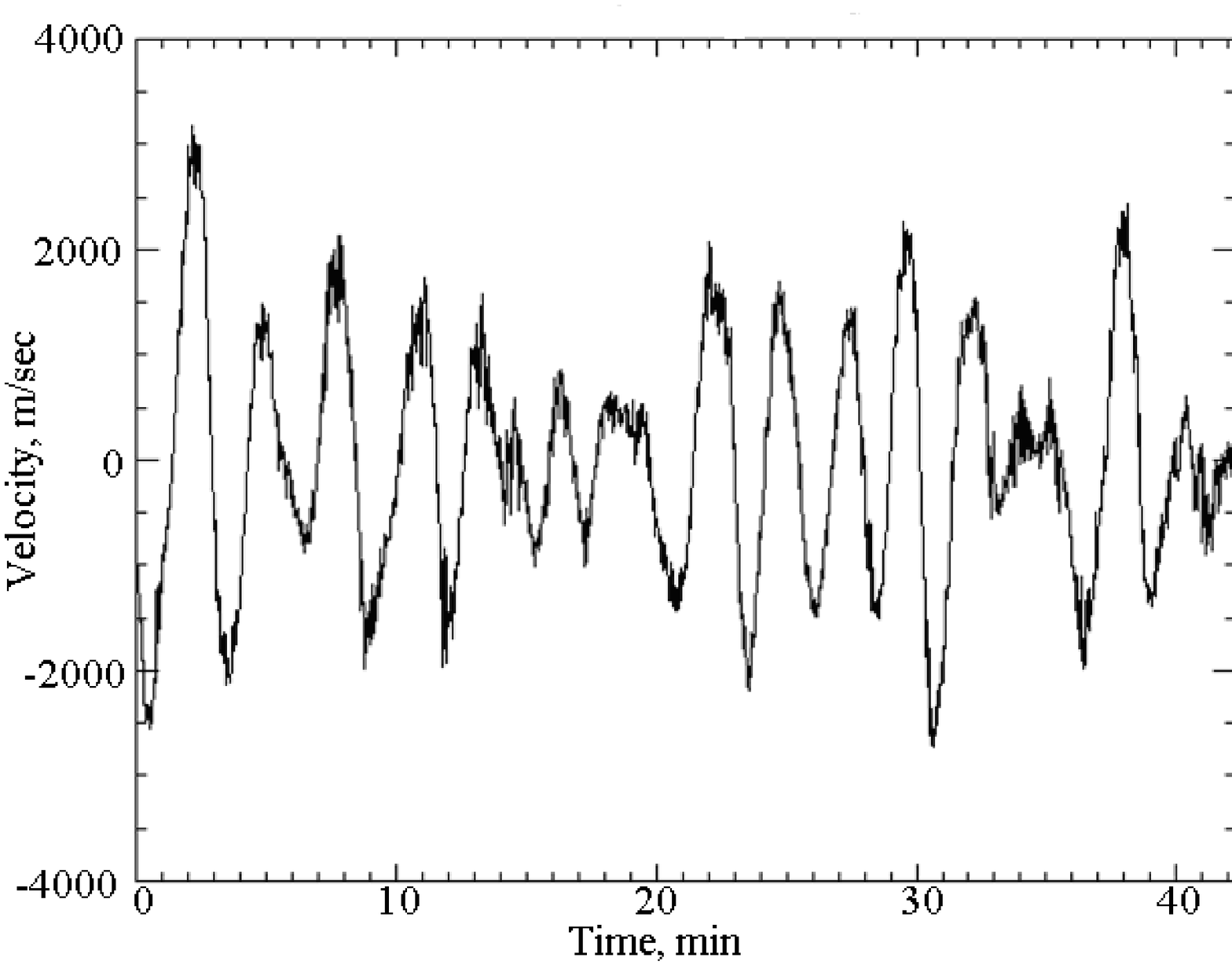}
               \includegraphics[width=0.515\textwidth,clip=]{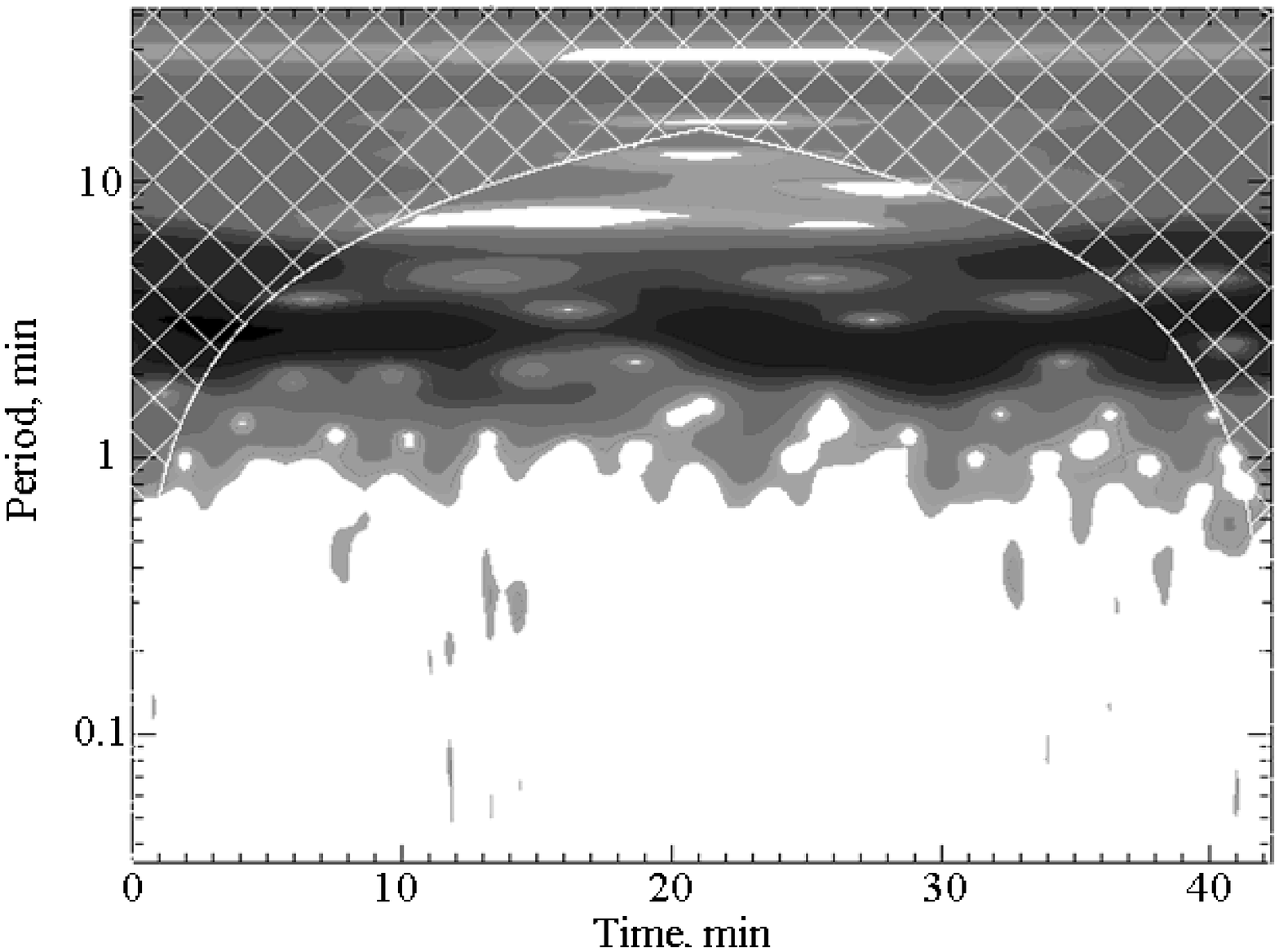}}
   \vspace{-0.35\textwidth}
   \centerline{\Large \bf
               \hspace{0.055\textwidth} \color{black}{c}
               \hspace{0.465\textwidth}  \color{white}{d}
   \hfill}
   \vspace{0.35\textwidth}
   \centerline{\includegraphics[width=0.515\textwidth,clip=]{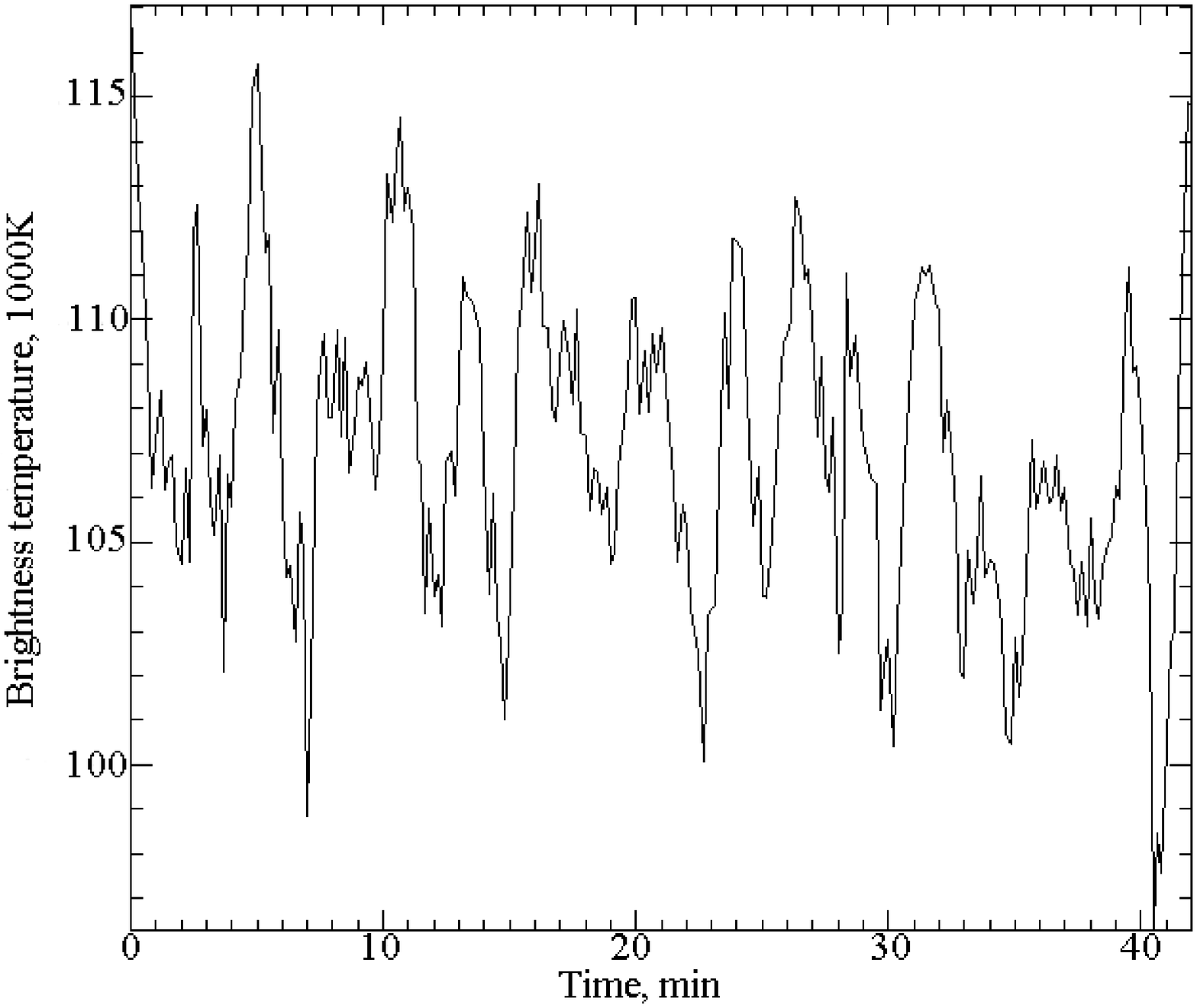}
               \includegraphics[width=0.503\textwidth,clip=]{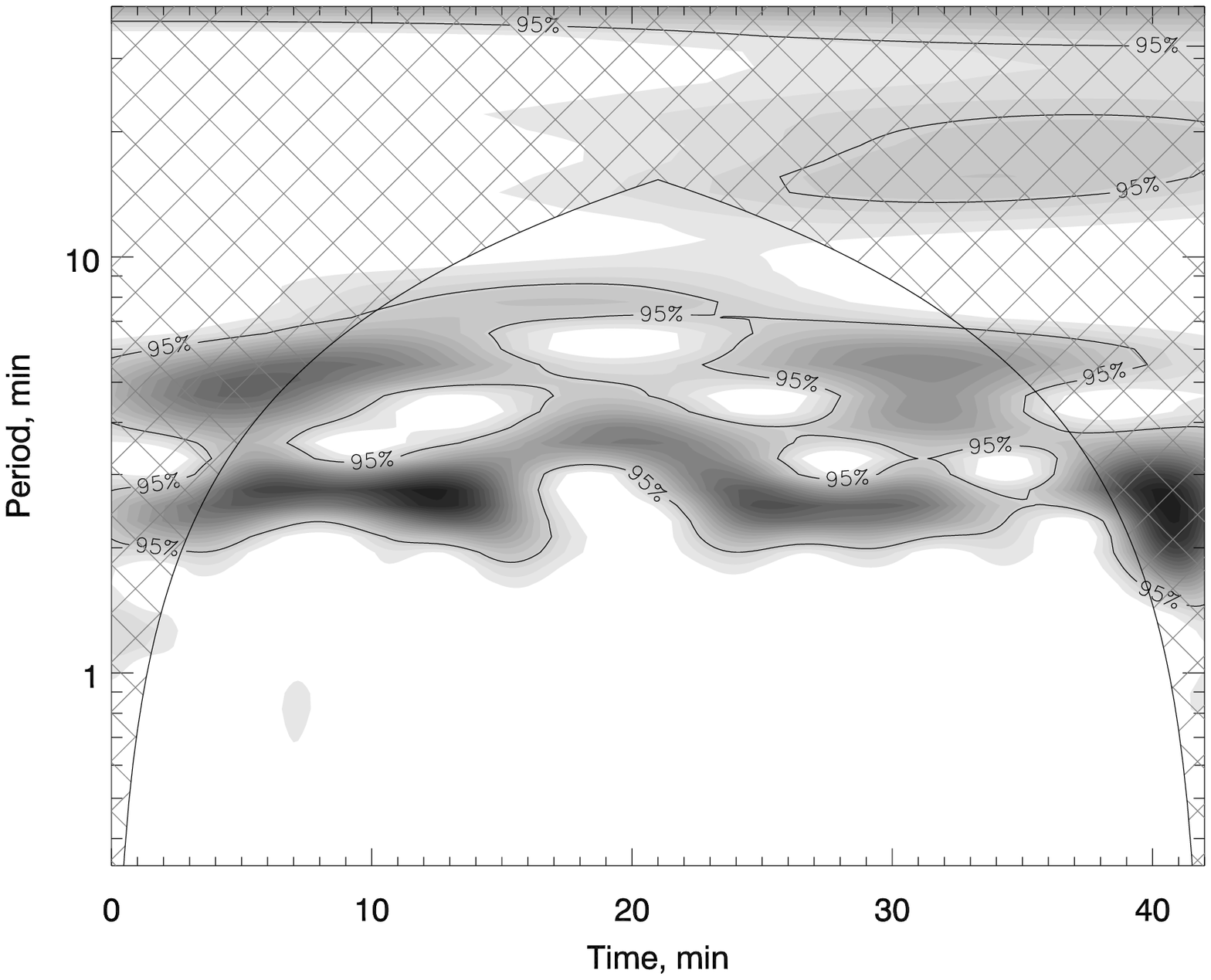}}
   \vspace{-0.4\textwidth}
   \centerline{\Large \bf
               \hspace{0.035\textwidth} \color{black}{e}
               \hspace{0.485\textwidth}  \color{black}{f}
   \hfill}
   \vspace{0.4\textwidth}
\caption{NOAA~661, 2004 August~18, 01:01\,--\,01:43 UT. Top: (a)
active region position on the solar disk (Mees Solar Observatory),
(b) radio map at 17~GHz. Middle: (d) the temporal variations and
(d) wavelet analysis of the chromospheric LOS velocity in
H$\alpha$. Bottom: (e) the temporal variations and (f) wavelet
analysis  of the radio emission of the bright source in AR661.
Darker regions correspond to higher power. The crosshatched area
shows the COI. The contours show the 95\% confidence level.}
\label{F-aug18}
\end{figure}

In spite of essential differences of the two methods of analysis,
the comparison of the spectra of the radio and optical
observations presents the possibility to study the appearance of
the same oscillation process at two levels in the solar
atmosphere. The identification of the oscillation mode is based on
a comparison of the frequency and amplitude variations revealed by
the wavelet transformation of the time series. In
Figures~\ref{F-aug15}--\,\ref{F-aug18} the samples of this
comparison are presented for three series of observations for
AR661~(2004~August~15, 16, and 18).

On 2004 August~18 the AR was situated near the center of the solar
disk (at approximately 0.2 solar radius, see
Figure~\ref{F-aug18}). Here one can easily see an apparent
similarity of the wavelet spectra obtained from optical and radio
observations. For the other two days this is not the case. The
difference is quite obvious and probably related to the position
of the AR on the disk. The point is that the chromospheric LOS
velocity can be easily measured when the AR is located close to
the solar center. Then it reflects the motion of a wave toward the
higher regions of the solar atmosphere. Therefore the radio
emission at the base of the solar corona is affected by this
motion. On the other hand, when the AR is far from the center of
the solar disk, optical methods are mostly relevant for the study
of oscillation modes across the magnetic tube of a sunspot, which
are not clearly visible at the CCTR measured by radio methods.
Some other possible reasons for the discrepancy will be discussed
later. One should also bear in mind that our sets of observations
lasted to 40\,--\,60 minutes, which is sufficiently long to study
the instability of the three minute oscillations, but provides
only limited possibilities for studying much longer periods.

For a number of cases our analysis confirmed that radio and
optical spectra of oscillations agree better when the AR is
located near the center of the solar disk. This peculiarity is
typical for the three-minute and five-minute oscillations. It
reflects MHD waves propagating along the magnetic tube of a
sunspot. When the sunspot is situated near the center of the solar
disk (the magnetic tube is directed toward the observer), we
expect the absence of the emission at harmonics of the
gyrofrequency. Accordingly there is no radiation at the very
center of the sunspot, but farther out there is radiation, which
is ring-shaped and turns into a horseshoe-shape away from the
center of the solar disk~\cite{Gelfreikh79,Nindos96}. The emission
is generated at the side of the sunspot's umbra where the magnetic
field strength may still reach 2000~G. These differences may be
caused by the above mentioned limitations in the radio image of
sunspots and the limitations of the optical data, which measure
only the longitude component of the velocity and only along the
slit of the spectrograph.

It is well known that most of oscillation processes observed in
sunspots are unstable with a low level of the quality of
oscillations. Both frequency and amplitude of oscillations vary
with time. Many of the observed wave trains are only of a few
periods length. To study the spectra, we used a wavelet transform.
At the same time this instability opens a new approach to compare
the parameters of the oscillations in two wavelength ranges,
comparing not only the frequency, but also the coherence.
Figure~\ref{F-aug18} gives a sample of the behavior. The identity
in the length of trains and their oscillation periods prove that
we register oscillations in the magnetic structure of a sunspot at
the levels of the chromosphere and corona.

\begin{figure}
\centerline{\hspace*{0.015\textwidth}
            \includegraphics[width=0.1655\textwidth,clip=]{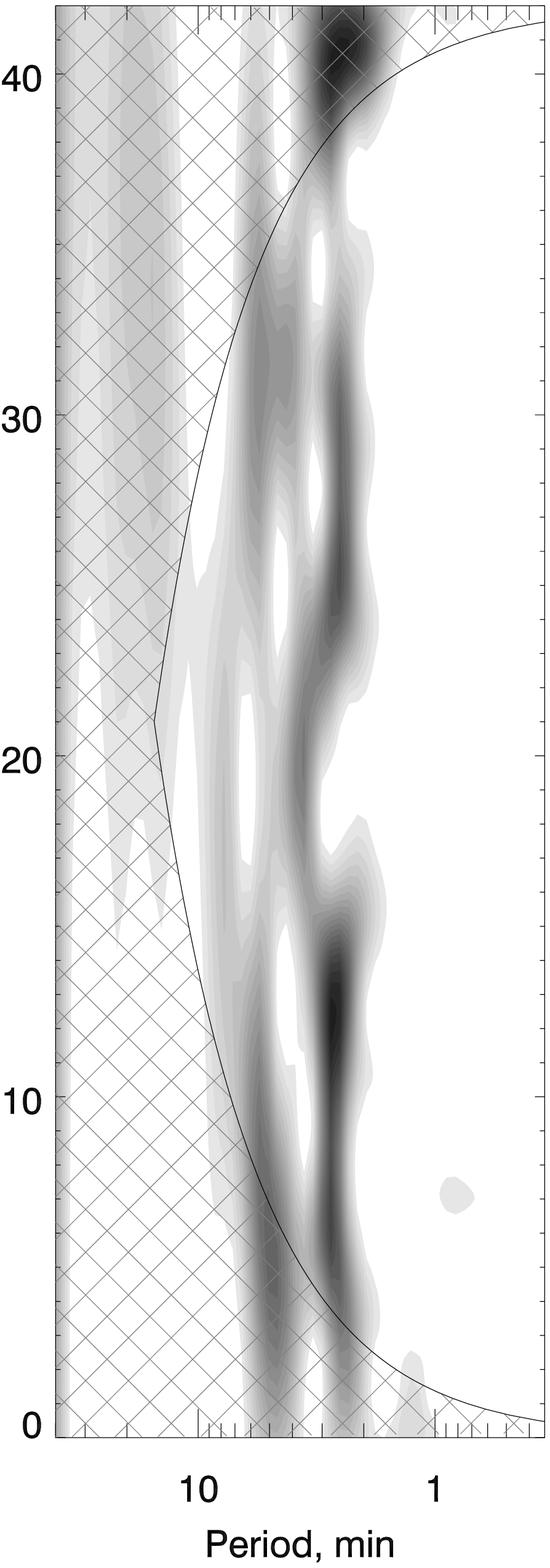}
            \raisebox{-5pt}{\includegraphics[width=0.4024\textwidth,clip=]{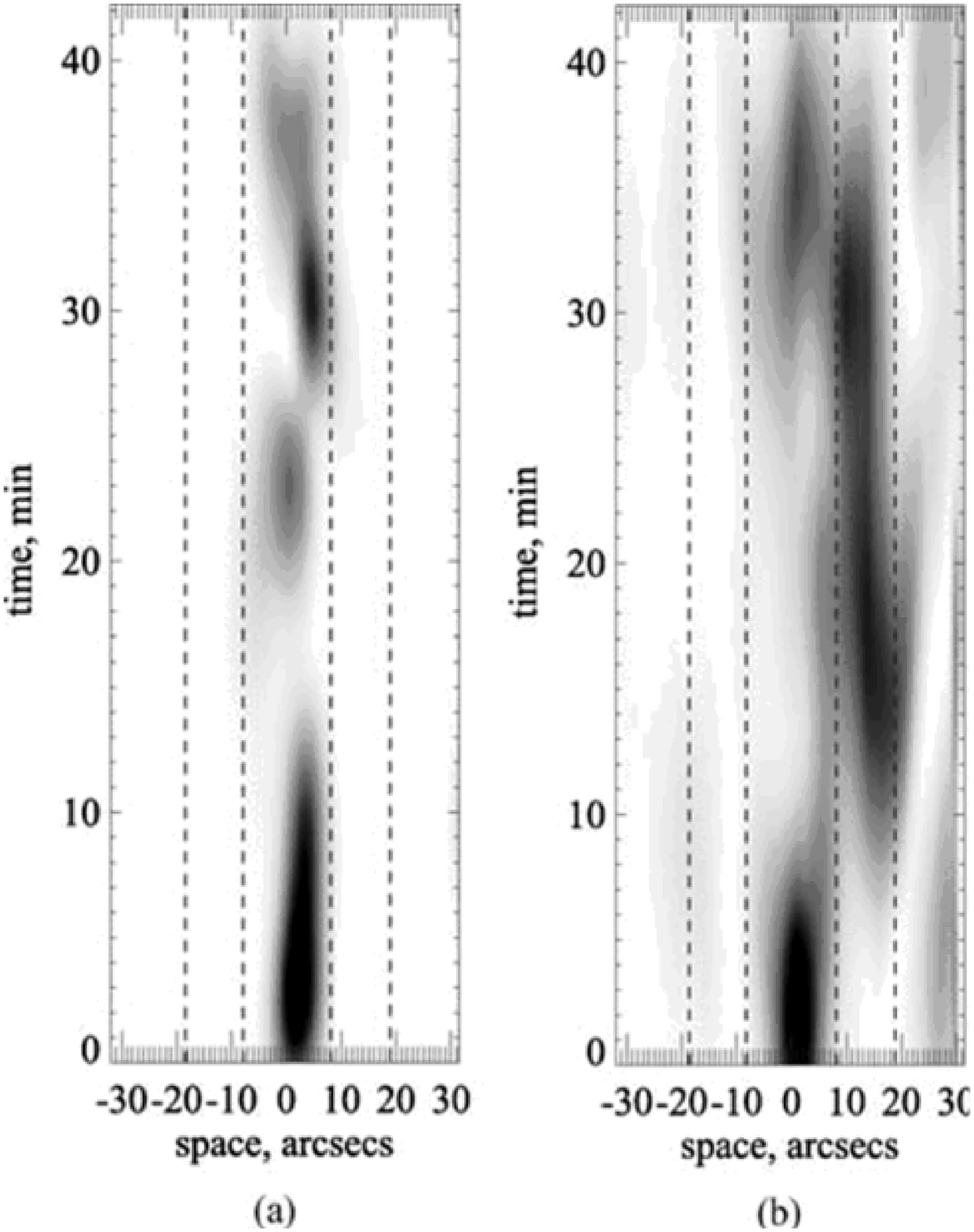}}
            \includegraphics[width=0.1552\textwidth,clip=]{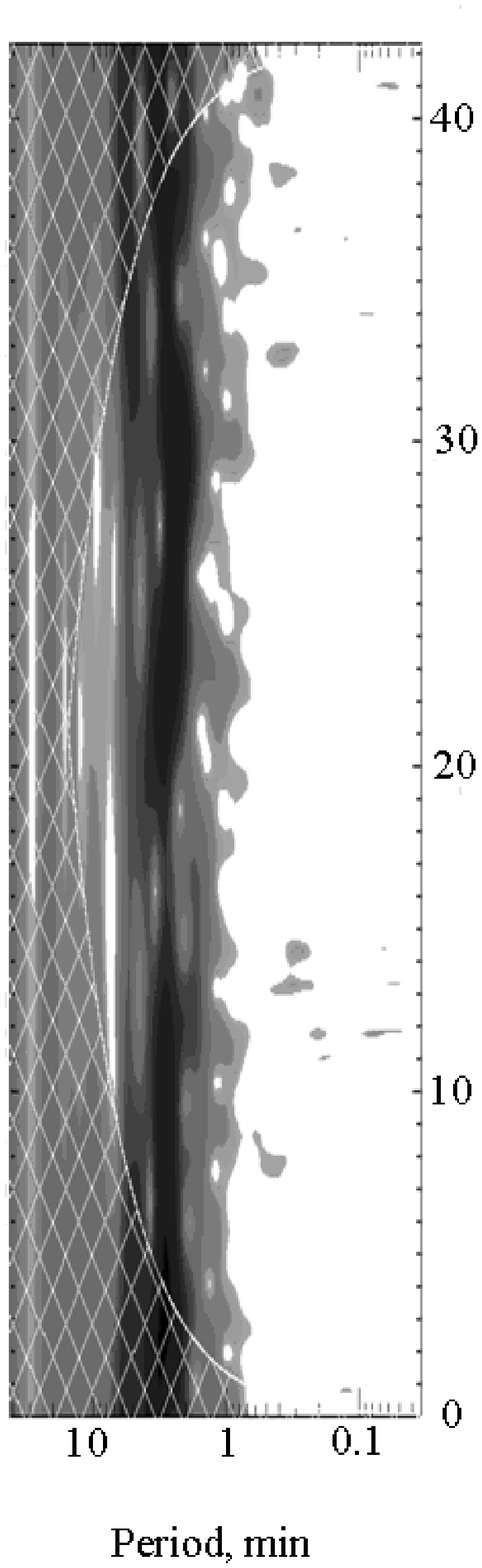}}
            \caption{NOAA~661,  2004 August~18, comparison of two types of
oscillations inside the spot. Space--time distribution of power (a
-- three-minute, b -- five-minute). Wavelet spectra for spot
umbra: left -- radio, right -- optical.} \label{F-comparison}
\end{figure}

\begin{figure}
   \centerline{\includegraphics[width=0.503\textwidth,clip=]{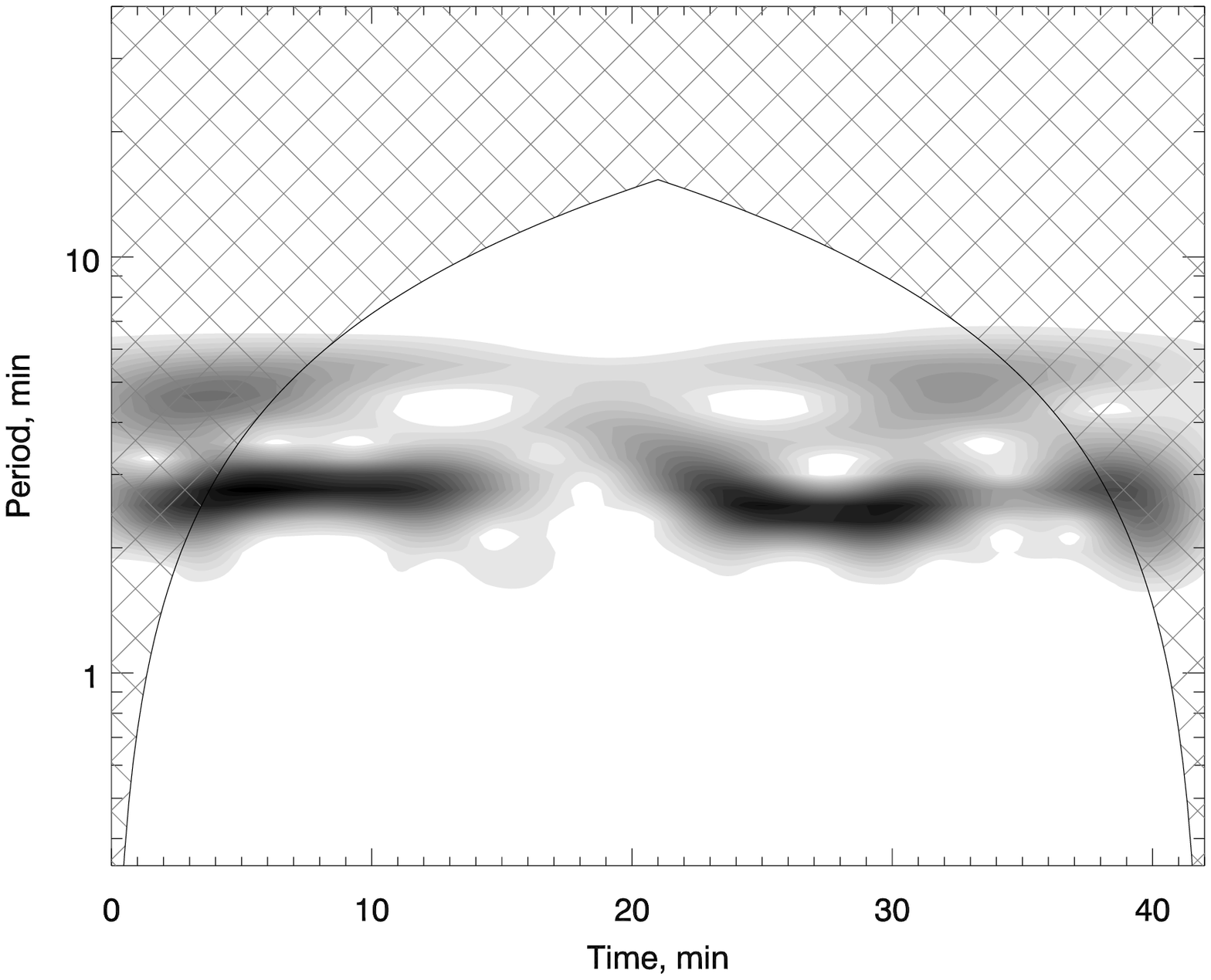}
               \includegraphics[width=0.503\textwidth,clip=]{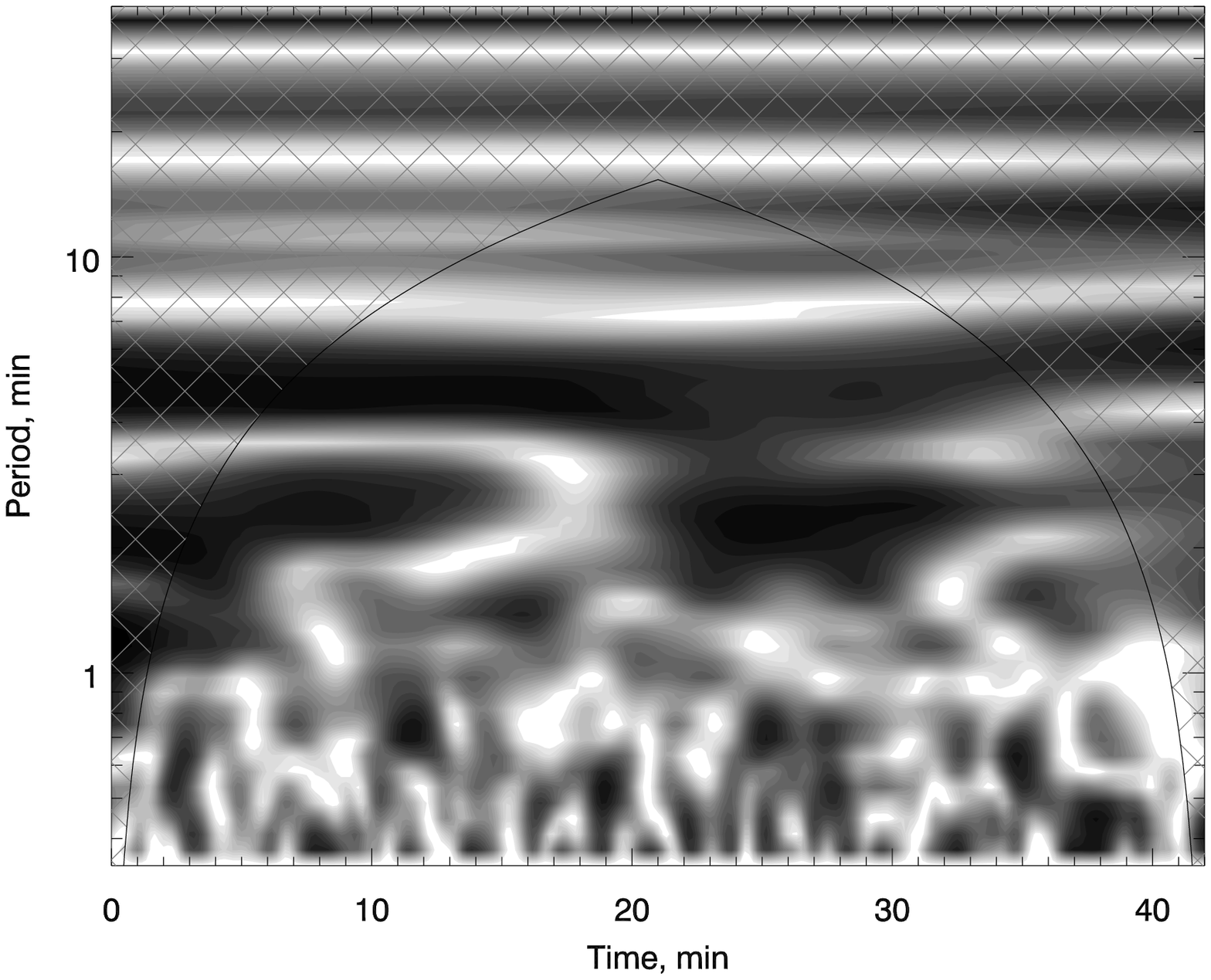}}
   \vspace{-0.4\textwidth}
   \centerline{\Large \bf
               \hspace{0.035\textwidth} \color{black}{a}
               \hspace{0.485\textwidth}  \color{white}{b}
   \hfill}
   \vspace{0.4\textwidth}
\caption{NOAA~661,  2004 August~18, 01:01\,--\,01:43 UT. (a)
Cross-wavelet transform between optical and radio time series. (b)
Wavelet coherency between optical and radio time series. Darker
regions correspond to higher power.} \label{cross_wavelet}
\end{figure}

\begin{figure}
   \centerline{\includegraphics[width=0.503\textwidth,clip=]{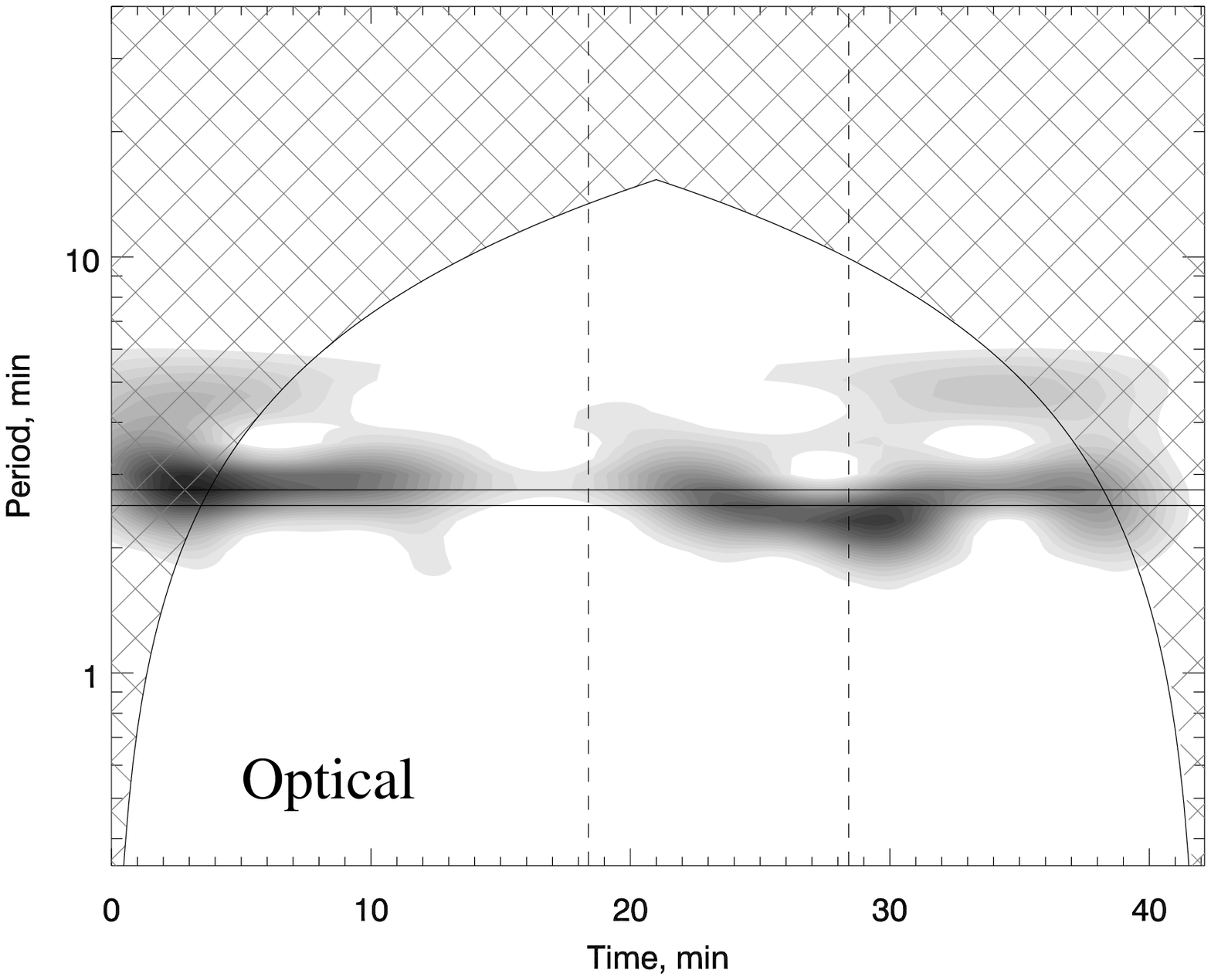}
               \includegraphics[width=0.503\textwidth,clip=]{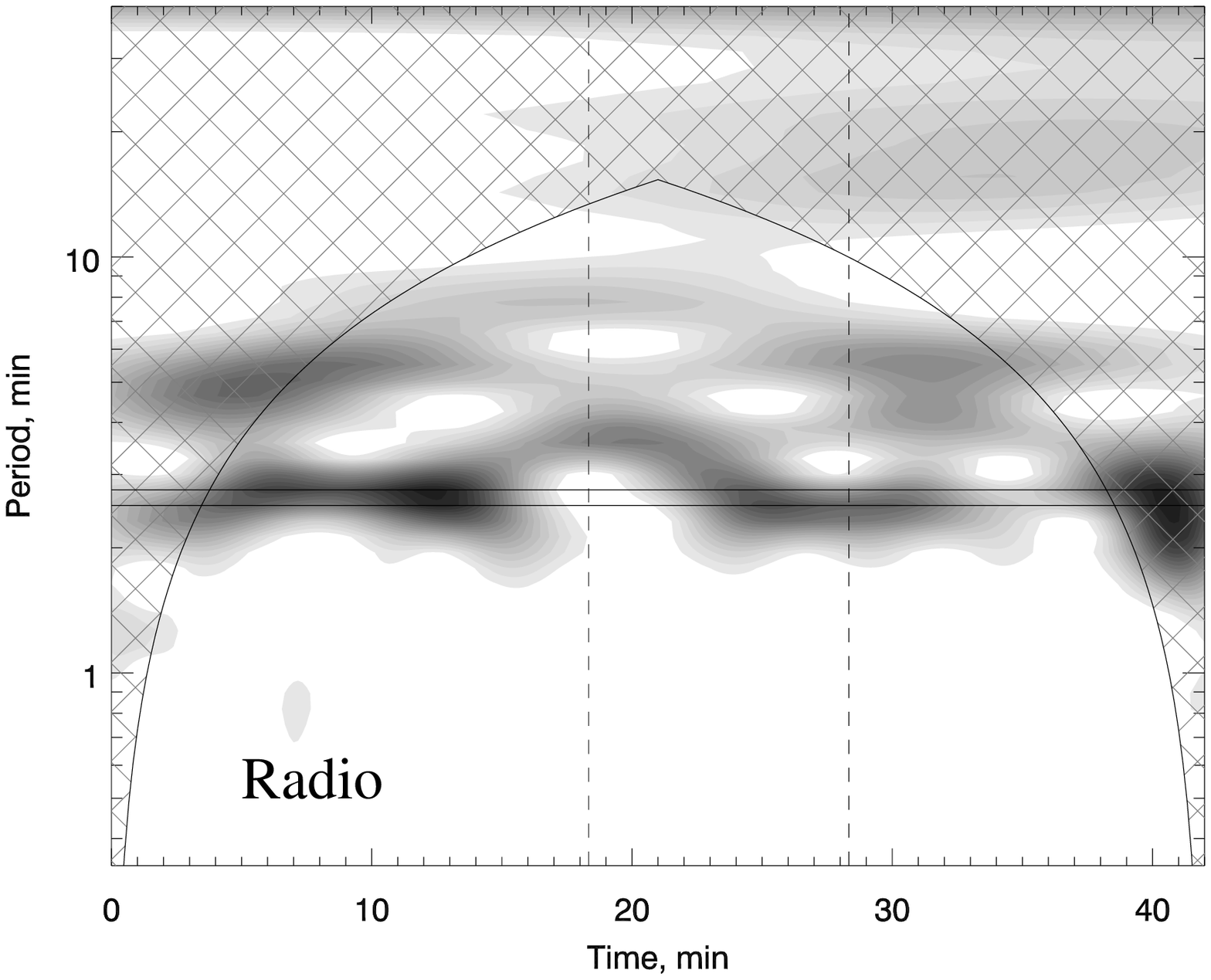}}
   \vspace{-0.4\textwidth}
   \centerline{\Large \bf
               \hspace{0.035\textwidth} \color{black}{a}
               \hspace{0.485\textwidth}  \color{black}{b}
   \hfill}
   \vspace{0.4\textwidth}
   \centerline{\includegraphics[width=0.503\textwidth,clip=]{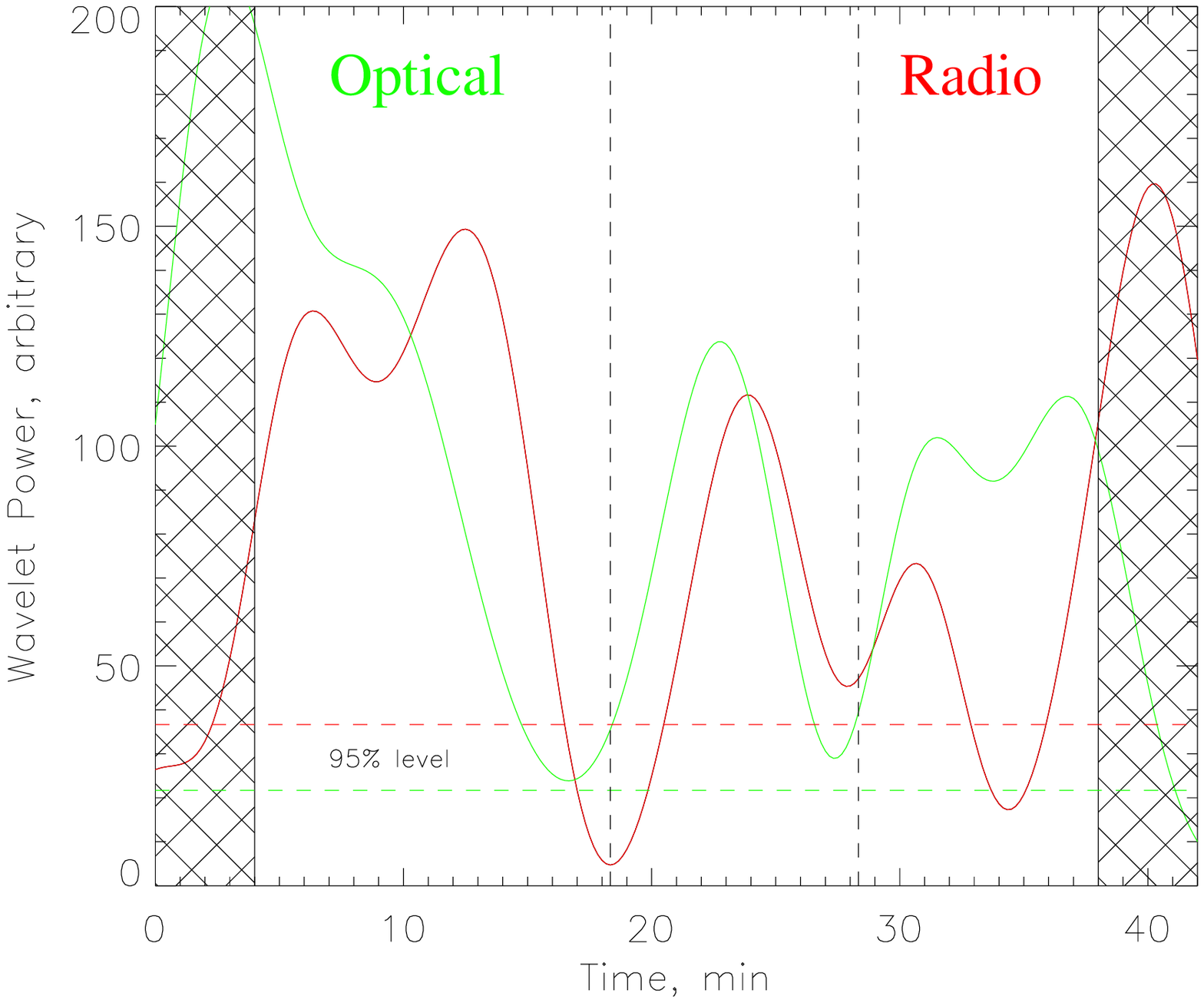}
               \includegraphics[width=0.503\textwidth,clip=]{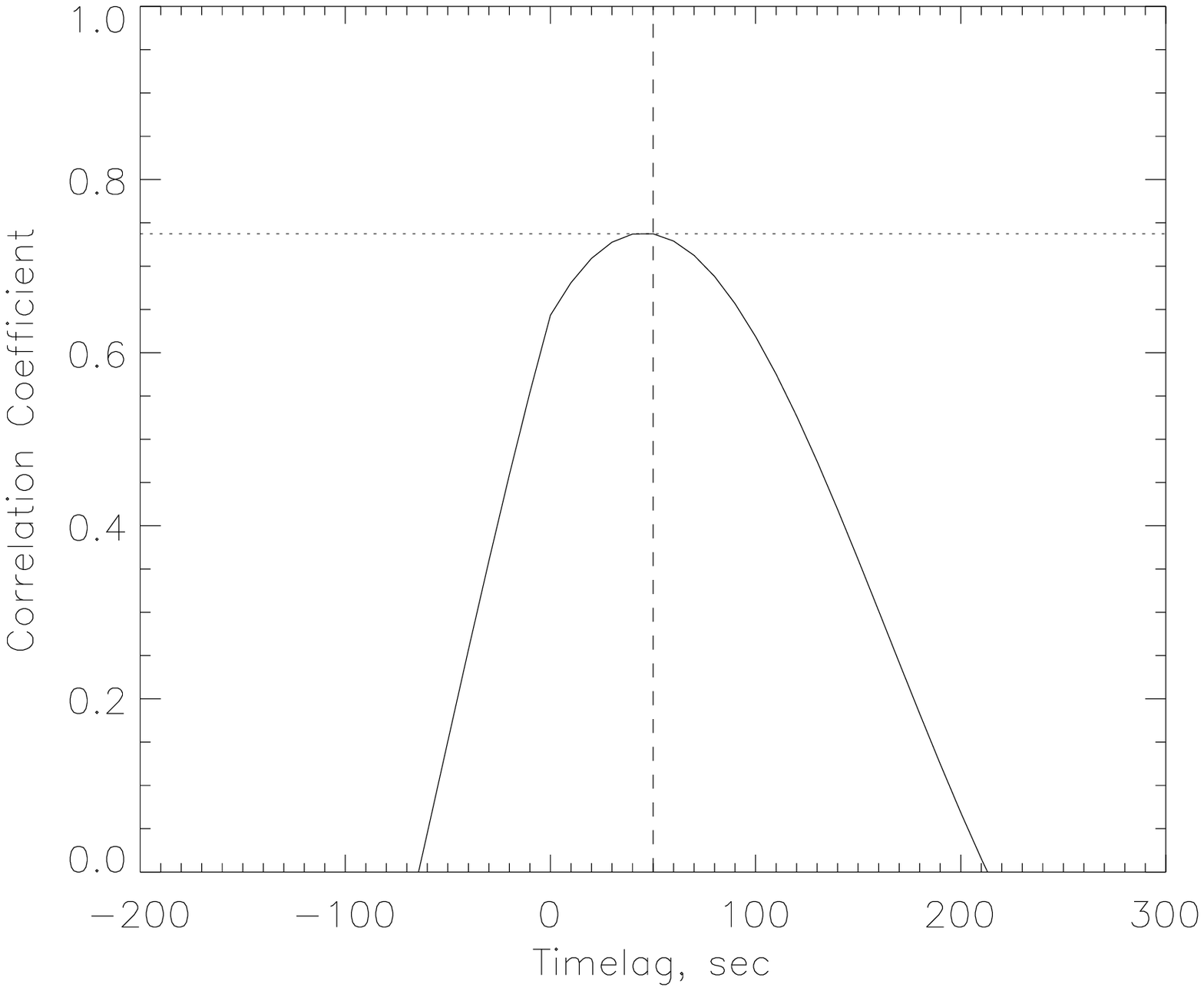}}
   \vspace{-0.4\textwidth}
   \centerline{\Large \bf
               \hspace{0.025\textwidth} \color{black}{c}
               \hspace{0.5\textwidth}  \color{black}{d}
   \hfill}
   \vspace{0.4\textwidth}
\caption{NOAA~661,  2004 August~18, 01:01\,--\,01:43 UT. Top: (a)
wavelet spectrum of the chromospheric LOS velocity in H$\alpha$,
(b) wavelet spectrum of the radio emission of the bright source in
the AR661. Bottom: (c) Power curves constructed by averaging the
wavelet power across the frequencies 6.0\,--\,6.5 mHz
(165\,--\,155 sec). The bands are shown in (a) and (b) by two
solid horizontal lines. Green curve shows the optical data, the
red curve shows the radio data. Horizontal dashed lines show  the
95\% confidence levels. (d) Cross-correlation curve between
fragment of curves shown in (c). The fragment is shown in (a),
(b), and (c) by dashed vertical lines.} \label{crosscorr}
\end{figure}

A detailed comparison of the wavelet radio and optical spectra
(Figures~\ref{F-comparison}\,--\,\ref{crosscorr}) shows two trains
of the three-minute oscillations with similar lengths and their
periods. Figure~\ref{cross_wavelet} depicts the cross-wavelet
transform $$W^{XY}_n(s)=W^{X}_n(s)\cdot{}W^{Y*}_n(s),$$ where
$W^{X}_n(s)$ and $W^{Y}_n(s)$ are wavelet transforms of time
series $X$ and $Y$, $n$ is the time, $s$ is the scale of the
wavelet, ``$*$'' signifies the complex conjugate, and wavelet
coherency~(Torrence and Compo, 1998; Torrence and Webster,
1999) %\cite{Torrence98,Torrence99}
between optical and radio time
series. Figure~\ref{crosscorr} shows the cross-correlation between
power curves constructed by averaging wavelet power across
frequencies 6.0\,--\,6.5 mHz (165\,--\,155~sec). To calculate the
cross-correlation, we used the method proposed
in~\opencite{Bloomfield04} (figure~2 on page~939). It is quite
evident that there is a shift in time between them. The radio
oscillations happen some $40-50$ seconds after the optical. This
result agrees well with the generally accepted idea that the
three-minute oscillations show an MHD wave spreading upward in the
corona~\cite{Lites92,Kobanov04a,Bogdan06}, its frequency is a
result of filtration in a lower photospheric region. In our case
the propagation velocity of these waves can be found from the
observations at the level of the chromosphere and is equal to
60~km\,{}s$^{-1}$~\cite{Kobanov06}. The actual velocity may be
even higher (bearing in mind the limitations of the method
employed). Therefore the delay of 45~sec is interpreted as the
time of propagation of the MHD wave from the chromosphere to the
corona  and suggests that the height of the CCTR is approximately
2700~km.

Concerning the horizontal phase velocity (60~km\,{}s$^{-1}$)
measured by us in a sunspot umbra, it is necessary to note that
observations by \textit{Hinode}~(Nagashima et al., 2007)
%\cite{Nagashima07}
have also given similar results (about
50~km\,{}s$^{-1}$). Within the  scope of the ``visual pattern''
scenario the acoustic waves traveling upward with a velocity of
about 10~km\,{}s$^{-1}$ can cause this
effect~(\opencite{Kobanov08}; see also~\opencite{Bloomfield07}).

Space--time variations of the oscillations of LOS velocity are
shown for three-minute oscillations (Figure~\ref{F-comparison}a)
and five-minute oscillations (Figure~\ref{F-comparison}b).
Variations of the spectra are also presented in
Figure~\ref{F-comparison} (Radio -- left and optical -- right).
The borders of the umbra and penumbra are marked by vertical lines
in Figures~\ref{F-comparison}a and b. The three-minute
oscillations present three trains; all of them are within the
umbra region, although in different positions. They are also
present in the wavelet spectra in both radio and optical data. For
five-minute oscillations we see five trains in optical picture,
but only some of them can be identified with the radio spectra.
The first in time train is definitely measured by the optical and
radio method. It covers most part of the umbra region. About
2\,--\,3 minutes delay is observed in the radio fluctuation. The
wave train at 30 minutes was identified in the radio spectra. In
the optical observations the oscillations are registered in the
penumbra region. This comparison is not obvious. The last
five-minute oscillation train is measured only in the center of
the sunspot. This region does not produce radio emission when
observed at the center of the solar disk, possibly resulting in
the observed peculiarity.

\section{Low-Frequency Oscillations in Sunspots}
\label{S-Low-frequency}

In contrast with the situation of the 3\,--\,5 minute
oscillations, longer periods of 8\,--\,30~minutes are rarely
reported. Nevertheless, one can confirm the presence of
long-periodic motions in the penumbra~\cite{Lites92,Brisken97} or
nearby~\cite{Kobanov00,Voort03}. It is worth mentioning that
because of the concentration of low-frequency modes of
oscillations at the outer regions of sunspots in optical
observations, we find an apparent decrease of the frequency of
running penumbral waves (RPW) from the center of a spot to
outside.

Rimmele (1995) and Kobanov and Makarchik (2003) %\inlinecite{Rimmele95} and \inlinecite{Kobanov03}
connected the 10\,--\,15~mi\-nutes oscillations with variations of
the Evershed velocity. Studies of oscillations with frequencies of
1\,--\,1.5~mHz (periods of 11\,--\,17~minutes) made from
space~\cite{Nagashima07} have shown the presence of oscillations
even at the inner region of a penumbra in the intensity of the
G~band. The active region was a single round sunspot. The
increased power of the above oscillations had the form of a ring
surrounding the umbra.

We had 42-minutes of radio data simultaneous with the optical
observations, which was long enough to find 8-minute and 16-minute
oscillations, although without a detailed analysis of their
stability. Long-period oscillations in the radio sources above
sunspots are already known~\cite{Gelfreikh06}.

\begin{figure}
   \centerline{\includegraphics[width=1.0\textwidth,clip=]{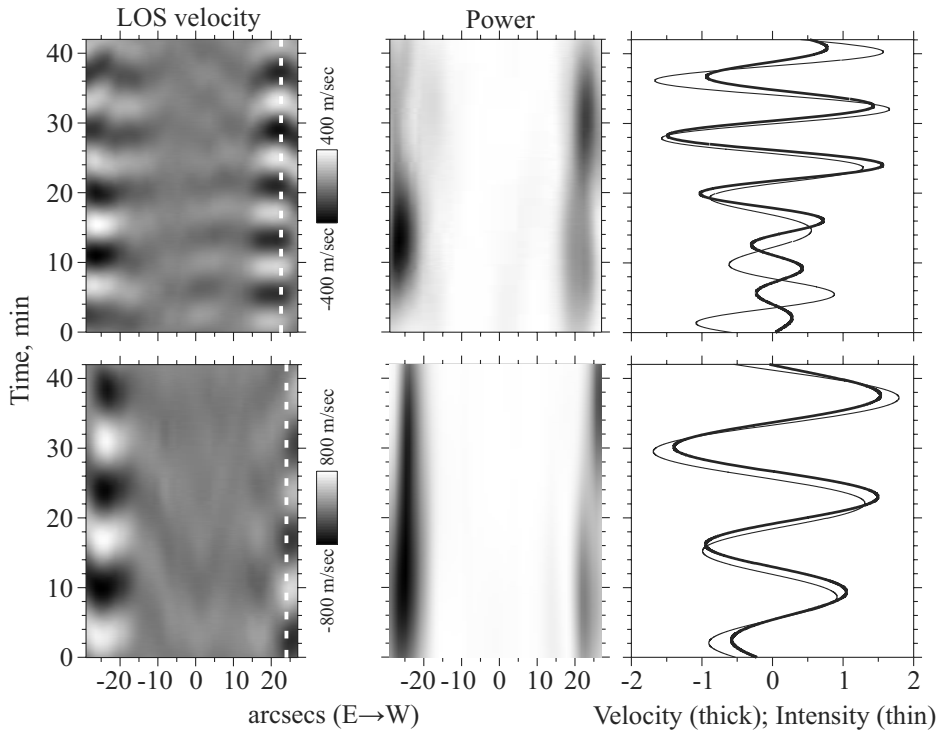}}
\caption{Chromospheric oscillations filtered for periods of 8.4
min (Top row) and 15 min (Bottom row). Left -- space--time
diagrams of LOS velocity. Middle -- the same for power of LOS
velocity oscillations. Right -- signals of LOS velocity and
intensity for the points marked by dashed line on the left
panels.} \label{F-diagram}
\end{figure}

Results of the optical observations, including the distribution of
oscillations across the spot, are illustrated in
Figure~\ref{F-diagram}. We can see that both the 8- and the
16-minute oscillations with a H$\alpha$ Doppler velocity are more
clearly visible outside the spot umbra, where the magnetic field
is inclined in the direction of the observer. However, the
strength of the field is evidently still powerful enough
($B\geq2000$~G) to produce radio emission. But because of the
insufficient duration of the optical series the estimate of the
time delay of the radio oscillations with respect to the optical
ones are hardly reliable.

Low-frequency oscillations (1\,--\,2~mHz) in the Doppler velocity
of the NOAA 661 were 22\,--\,25~arcsec from the center of the
spot, at the east and west directions along the slit of the
spectrograph, measured at the outer side of the penumbra of the
sunspot. Figure~\ref{F-diagram} presents the space--time diagrams
of the Doppler velocity along the slit for the periods of 8.4 and
15~minutes. For the 15-minute oscillations the declination of the
strips in the diagrams shows the propagating waves. In the eastern
and western parts of the region the horizontal component of the
motion is directed from the east to the west. For the 8.4~minute
oscillations diagram the east part of the oscillations shows less
inclined strips toward the horizontal direction than those in the
west position, possibly because of standing waves. For the
eight-minute oscillations the direction of horizontal propagation
in the western part is inverse to the eastern one and coincides
with the direction of Evershed effect in the chromosphere. The
phase velocity estimate (horizontal component) gave the value of
30\,--\,50 km\,s$^{-1}$. It is interesting that the intensity
oscillations are practically in phase with the Doppler velocity
oscillations. It is common knowledge that the phase lag between
velocity and intensity is a quarter-period for the acoustic
oscillations.

\begin{figure}
   \centerline{\includegraphics[width=1.0\textwidth,clip=]{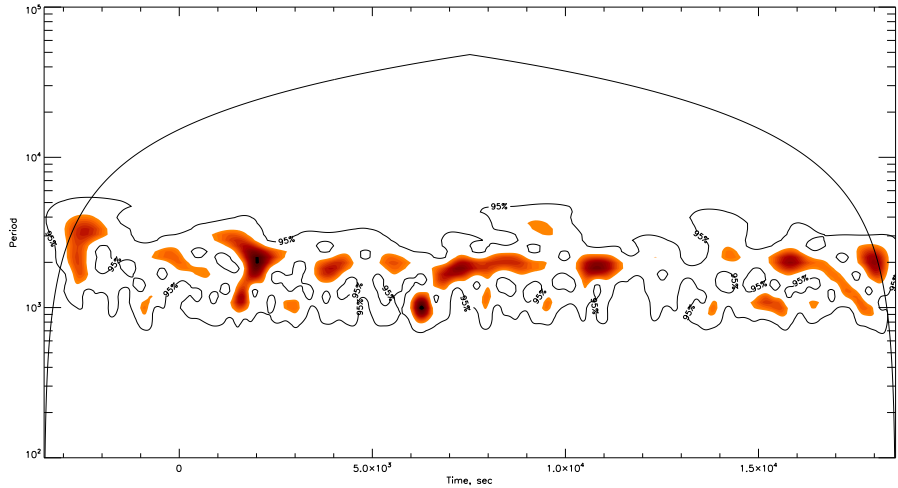}}
\caption{Wavelet spectrum of the radio sunspot-associated source
in NOAA661 on 2004 August~18 at wavelength $\lambda=1.76$~cm. }
\label{F-wavelet}
\end{figure}

\begin{figure}
   \centerline{\includegraphics[width=1.0\textwidth,clip=]{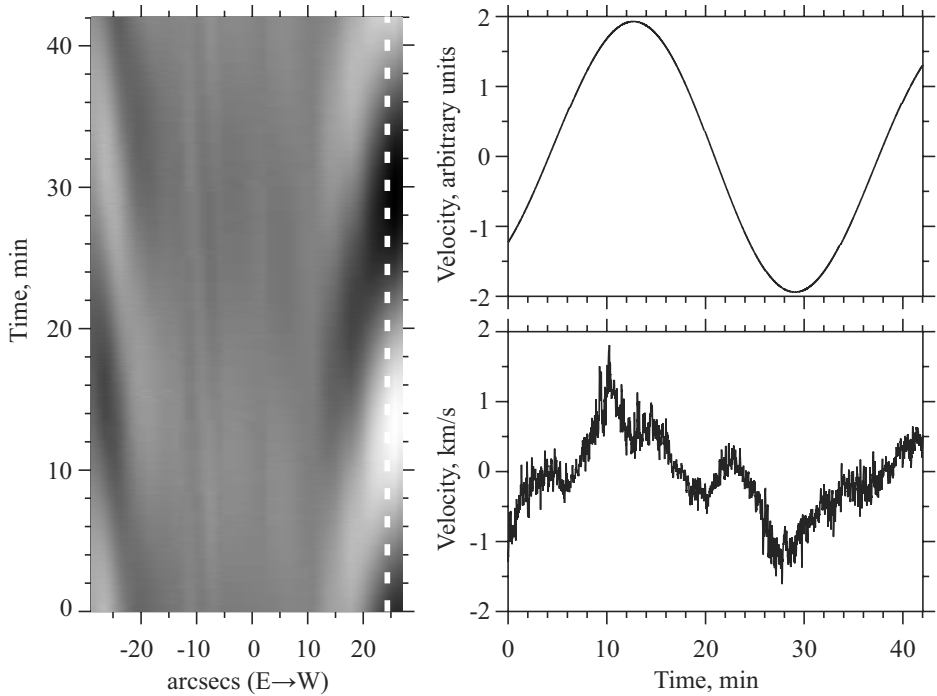}}
\caption{Example of 32-minute oscillations in NOAA 661. Left:
filtered space--time diagram, Right: LOS velocity signal in point
24\,--\,25 arcsec -- filtered (top) and unfiltered (bottom).}
\label{F-long}
\end{figure}

However, to obtain more conclusive information on the longer
period oscillations we needed to analyze a longer period of
observations, which was available at the radio observations at
Nobeyama. The wavelet analysis of these observations is presented
in Figure~\ref{F-wavelet}. In these spectra one can clearly see
the oscillations with periods of 15\,--\,16 and 30\,--\,32
minutes. Both modes are essentially non-stationary with minor
variations of the amplitude and the period. An essential
peculiarity of the quasi-periodic process is the strong similarity
of the variations of parameters for both modes. The 30-minute
oscillations happen with a delay of several minutes with respect
to the 15-minute oscillations. Although the ratio of the
frequencies of the two modes is not exactly equal to~2
(about~1.9), the similarity of the temporal variations of the two
modes lead us to the conclusion that the 16-minute oscillations
are the harmonic of the 30\,--\,32-minute oscillations. The delay
in oscillations of the 30-minute mode may be the result of the
propagation dispersion of the modes from the region of their
simultaneous generation (owing to the difference in phase
velocities of the MHD waves). The duration of the analyzed optical
observations does not allow us to study the 32-minute wave.
Nevertheless, the oscillations with this typical temporal
variation is clearly registered (see Figure~\ref{F-long}).

\section{Discussion}
\label{S-Discussion}

The above analysis of quasi-periodic oscillations measured at the
chromospheric level in the H$\alpha$ line and radio confirms the
oscillations with periods from 160 to 200~seconds above sunspots
in the lower corona~\cite{Brynildsen02}. They are certainly the
counterpart of the three-minute oscillations studied by optical
methods for several decades~\cite{Lites92}. These oscillations at
the level of the chromosphere observed in the H$\alpha$ line were
clearly shown. Their theory was developed by many
authors~\cite{Zhugzhda82,Staude99,Zhugzhda00}. First they were
considered to be oscillations of a magnetic tube resonator of
sunspots limited by the CCTR and some surface below. Later
developments in observations and in theory related them to
propagating MHD waves that were filtered in frequency in some
lower under-photospheric layer. The solution of the problem is far
from clear.

To our knowledge, we present here for the first time the study of
these MHD waves based on simultaneous optical and radio
observations, which refer to coronal and chromospheric levels of
the solar atmosphere. We investigated the sunspot oscillations at
different levels of the solar atmosphere with a wavelet analysis
for the physical type of oscillations measured in radio and
optical wavelengths. We expected and measured some time delays of
the radio emission resulting from the sunspots assumed propagation
of the disturbance (MHD waves, probably) from the chromosphere to
corona. These time delays were found to be about 50~seconds for
the three-minute oscillations and a bit longer (about three
minutes) for the five-minute oscillations.

In all cases of our analysis the position of the AR near the
center of the solar disk resulted in a higher degree of similarity
of the observed oscillation parameters found from the radio and
optical methods. This situation looks natural if one takes into
account that we compared velocity oscillations at photospheric and
chromospheric levels (optical data) with oscillations at the CCTR
or lower corona (radio observations). In optical observations the
oscillations are more effectively seen at the longitude component
of the velocity. When the AR is near the center of the disk, the
optical observations show waves directed toward the upper layers
of the solar atmosphere, where the registered radio emission above
the sunspot is generated. However, in this case the central part
of the magnetic tube of the sunspot is beyond the scope of the
analysis because no emission at the harmonics of the
gyro-frequency is then generated. At the edge of the umbra the
magnetic fields are weaker and radio emission is generated at
lower heights, possibly with a higher temperature gradient. Then
the radio source normally has a ring structure reflecting the
upper limit of the umbra region~\cite{Gelfreikh79}. For a radio
source at a significant distance from the center of the solar disk
the radio emission may be measured at the stronger central part of
the sunspot umbra with a higher source temperature of the radio
emission, but the optical observation of the LOS velocity presents
effects mostly in a different direction of the wave propagation.

One important reason for the discrepancy of the radio and optical
analysis of the observations is connected with the fact that in
the optical data we analyzed only a one-dimensional scan of the
spot, while in radio it was the sum of the radio emission along a
ring or horseshoe structure. One cannot expect a strong symmetry
in the oscillation process (see also some results shown above).
The result is certainly an important factor for the differences
between wavelet spectra in radio and optical observations. Besides
the reasons of a methodical nature discussed above, the real
differences in the oscillation processes and their spectra could
be essentially caused by differences in plasma and magnetic
structures. That different sunspots may show quite different
spectra of oscillations, was found and illustrated before at the
same wavelength~\cite{Gelfreikh06}.

Longer oscillations in sun\-spots are also known, though in a much
lower stage of analysis. Therefore we also explored this
direction. For the radio data we could use observational data of
the Nobeyama Radioheliograph of several hours length. In this case
we detected clearly visible periods of oscillations, though of low
coherence, slightly longer than 15 and 30~minutes. The obvious
similarity of variations of amplitude and periods lead us to the
conclusion about the common origin of these modes. The 15-minute
mode may possibly be the harmonic of the longer one. In the
present case we had only 42~minutes of observations of the same
sunspot in the optical and radio ranges. The 15\,--\,16~minute
oscillations are well measured. The 32-minute variation was also
found. A detailed similarity of these oscillations at the
chromosphere (optical) and the corona (radio) is impossible to
determine from our data, in contrast to the results of our
analysis of the three- and five-minute oscillations for the same
sunspot. The 15-minute variations in the optical measurements are
stronger in the penumbra region, although they are also present in
the center of the umbra. Further studies with a longer duration of
the optical measurements should be organized to solve the above
uncertainties of the data comparisons. As the eight-minute
oscillations in the chromosphere, their absence in the radio
spectra possibly requires further analysis with special filtration
to achieve higher sensitivity.

\section{Conclusion}
\label{S-Conclusion}

This work illustrates the recent progress in a new approach of
studying oscillations of active regions, and sunspots especially,
by comparing the parameters of the observed oscillations at
different levels of the solar atmosphere. The value of the radio
astronomical observations made with the Nobeyama Radioheliograph
enables us to measure the temporal variations of the plasma
parameters inside the sunspot magnetic tube at the basis of the
corona. The regular daily observations cover more than 15~years
with practically any desirable cadence. The comparison of these
observations with those of the chromospheric velocity in H$\alpha$
resulted in some new estimates of the 3D structures and dynamics
of the oscillations in a sunspot.

It is generally accepted that MHD waves play an important role in
the transfer and release of nonthermal energy in the solar
atmosphere, therefore the QPOs present an interesting approach in
analyzing the parameters of the plasma structures and their
stability in the solar atmosphere. Knowledge of their properties
is essential in the development of realistic theories of such
fundamental processes as heating the corona, an accumulation and
release of the energy in flares, CMEs, and similar processes,
typical for our Sun and other stars. We showed the usefulness of
developing and applying the necessary efforts in similar
directions, covering the wide range of analyzed periods of
oscillations and longer time intervals of analysis. The physical
interpretation of the results requires more theoretical and
modeling analysis.

%%%%%%%%%%%%%%%%%%%%%%%%%%%%%%%%%%%%%%%%%%%%%%%%%%%%%%%%%%%%%%%%%%%%%%%%%%%
\begin{acks}
This work is supported by RFBR grants 08--02--91860-RS-a,
06--02--16838 and 10-02-00153-a, and also grants by the Program of
Support for Scientific Schools of the Russian Federation
(NSh--6110.2008.2 and Nsh--2258.2008.2) and the Federal Agency for
Science and Innovation (State Contract 02.740.11.0576). We are
grateful to the anonymous referee for important remarks, which
helped us to improve the paper.

Data used here from Mees Solar Observatory, University of Hawaii,
are produced with the support of NASA grant NNG06GE13G.

Wavelet software was provided by C.~Torrence and G.~Compo, and is
available at URL: http://paos.colorado.edu/research/wavelets/

\end{acks}

\end{article}

\end{document}